\documentclass[12pt]{article}
\usepackage{amsmath,amsfonts, epsf,epsfig,color,latexsym,mathrsfs}
\usepackage{bbold}
\numberwithin{equation}{section}

\ifx\ltimes\undefined
\newcommand{\ltimes}{{\kern3pt\hbox{\vrule width 0.4pt height 5.30pt
depth .0pt}\kern-1.76pt\times\kern1pt}} \fi

\ifx\lrtimes\undefined
\newcommand{\rtimes}{{\kern1pt\times\kern-4.76pt\kern3pt\hbox{\vrule width 0.4pt height 5.30pt
depth .0pt}}} \fi

\def\Z {\mathbb{Z}}
\def\R {\mathbb{R}}

\def\ti{\tilde}

\def\bid{\hbox{1\hspace{-0.04in}I}} %blackboard bold 1

\def\d{\text{d}}
\def\G{\Gamma}

\def\cG{{\cal G}}
\def\cX{{\cal X}}

\begin{document}

\begin{titlepage}

\vspace*{40mm}

\begin{center}

\textbf{\Large {Bi-algebras, generalised geometry and T-duality}} \\

\vspace*{15mm}

{R A Reid-Edwards} \\
\vspace*{10mm}

{\em Centre for Mathematical Sciences }\\
{\em City University, London } \\ {\em Northampton Square} \\
{\em London, EC1V 0HB, U.K.} \\

\vspace*{20mm}

\end{center}

\begin{abstract}

A study of sigma models whose target space is a group $G$ that admits a compatible Poisson structure is presented. The natural action of $O(D,D;\Z)$ on the generalised tangent bundle $TG\oplus T^*G$ and a generalisation of the Courant bracket that appears are reviewed. This background provides a concrete example where the generalised geometry and doubled geometry descriptions are both well understood. Connections between the two formalisms are discussed and the world-sheet theory from Hamiltonian and Lagrangian perspectives is investigated. The comparisons between the approaches given by generalised geometry and doubled geometry suggest possible ways of generalising the analysis beyond the known examples.

\end{abstract}

\vfill

\noindent {Ron.Reid-Edwards.1@city.ac.uk}

\end{titlepage}

\newpage

\section{Introduction}

The starting point of Buscher's proof of T-duality in \cite{Buscher ``A Symmetry of the String Background Field Equations''} was to consider a
background in which $D$ of the dimensions are compact. Let the metric, dilaton and $H$-field along the compact directions are denoted by
$g_{ij}$, $\phi$ and $H=\d B$ respectively. The compact space is then required to have an abelian isometry group generated by a set of isometry
vectors $K_m=K_m{}^i\partial_i$ such that
\begin{equation}\label{isometry}
\mathcal{L}_Kg_{ij}=0   \qquad  \mathcal{L}_KH=\d\imath_KH=0    \qquad  \mathcal{L}_K\phi =\imath_K\d\phi=0
\end{equation}
The canonical example of such a background is a torus fibration. In order to demonstrate the duality, the invariance is elevated
to a gauge symmetry on the world-sheet and one-forms $A^m$ are introduced, transforming as connections for the abelian gauge theory. Introducing
a Lagrange multiplier term $\d\tilde{x}_m\wedge A^m$ into the gauged theory one can show that, not only are the connections flat but, if the
compact space has the correct periodicity conditions, that the holonomies of the connections also vanish\footnote{The vanishing of all
holonomies of a flat connection is enough to ensure that  we can fix $A^m=0$.}. Thus the gauged theory is equivalent to the un-gauged theory. Alternatively
one may integrate out the gauge fields $A^m$ in the gauged sigma model and re-express the gauged theory in terms of the Lagrange multipliers $\tilde{x}_m$. This dual
formulation of the theory is equivalent to the gauged world-sheet theory with one-forms $A^m$ and therefore equivalent to the original, un-gauged theory we first thought of. The background fields of
the two dual formulations are related by the action of $O(D,D;\Z)\subset O(D,D)$ \cite{Giveon:Generalized duality in curved string backgrounds}.
The construction was generalised to include Ramond-Ramond fields in \cite{Hassan:1999bv,Hassan:2000kr} and for non-trivial torus fibrations in
\cite{Hull ``Global Aspects of T-Duality Gauged Sigma Models and T-Folds''}. A comprehensive review of the early research in this area, including many applications, may be found in \cite{Giveon:Target
space duality in string theory}.

Following from the success of the Buscher construction many authors\footnote{See \cite{de la Ossa:1992vc} in particular} attempted to generalise
T-duality to backgrounds in  which the isometry group is non-abelian. As in the abelian case, world-sheet fields $A^m$ are introduced to gauge the
non-abelian isometry group and a Lagrange multiplier term $tr(\tilde{x}_mF^m)$ is introduced to constrain the non-abelian field strength to
vanish. A potential problem was highlighted in the fact that the `dual' theory generally will not have isometries so it is unclear how one would
demonstrate the duality in the other direction. In fact the problems constructing a non-abelian generalization of the Buscher rules run far
deeper. It was shown in \cite{Giveon:1993ai} that the pairs of backgrounds produced in this non-abelian construction are generally not dual.
Though it is possible to ensure the connection is flat ($F^m=0$), it is not possible to remove the holonomies of $A^m$, so that the gauged theory is not
physically equivalent to the original theory. At best, the non-abelian `duality' is map between inequivalent sigma models. At tree level this non-abelian `duality' is a genuine symmetry as the world-sheet is topologically a sphere and all holonomies are therefore trivial. This fact has been exploited recently \cite{Berkovits:2008ic} to construct a fermionic version of T-duality to explain the presence of a dual superconformal symmetry in colour-stripped planar scattering amplitudes of ${\cal N}=4$ super Yang-Mills.

In \cite{Klimcik:1994gi,Klimcik:1995dy,Klimcik:1995ux} the notion of `Poisson-Lie' duality was introduced in which the non-abelian
map\footnote{The term `duality' will be reserved only for those maps which relate physically equivalent quantum sigma models, i.e. symmetries of the
string theory.} described above was generalised to include backgrounds without isometries on \emph{both} sides of the map. The background need not satisfy the isometry conditions
(\ref{isometry}) but must satisfy the much weaker (`non-isometric') condition
\begin{equation}\label{poisson lie condition}
\mathcal{L}_m\mathcal{E}_{ij}=c_m{}^{np}K_n{}^kK_p{}^l\mathcal{E}_{ik}\mathcal{E}_{lj}  \qquad  {\cal L}_m\phi=0
\end{equation}
where $\mathcal{E}_{ij}=g_{ij}+B_{ij}$ is the background field and $K_m=K_m{}^i\partial_i$ are vector fields which satisfy the commutator
$[K_m,K_n]=f_{mn}{}^pK_p$ where $f_{mn}{}^p$ are structure constants for a non-abelian group. The $c_m{}^{np}$ in (\ref{poisson lie condition}) are also structure constants such that there is a $2D$ dimensional Lie-algebra, called a Drinfel'd double
with commutation relations
\begin{equation}\label{**}
[T_m,T_n]=f_{mn}{}^pT_p \qquad [T_m,\widetilde{T}^n]=c^{np}{}_mT_p-f_{mp}{}^n\widetilde{T}^p \qquad
[\widetilde{T}^m,\widetilde{T}^n]=c^{mn}{}_p\widetilde{T}^p
\end{equation}
The details of the derivation of this condition can found in \cite{Klimcik:1995dy}. Despite the elucidation that this approach gives to the
non-abelian map, it still falls foul of the arguments presented in \cite{Giveon:1993ai} and in general does not describe a duality of the string
theory. Recent studies of gauged supergravities have lead to a renewed interest in T-duality on non-isometric backgrounds. Though it is clear
that the Poisson-Lie map as it stands is not \emph{generally} a true duality of the string theory, there are examples in which the map is a
duality\footnote{The abelian Drinfel'd double $\cG=U(1)^{2D}$, corresponding to a compactification on $T^D$, is an obvious example.}
and leads to interesting string theory backgrounds.

\subsection{Gauged supergravity and T-duality}

The duality discovered by Buscher is manifest in the effective supergravity theory as part of a rigid $O(D,D;\Z)$ symmetry
\cite{Giveon:Generalized duality in curved string backgrounds}. The classical supergravity equations of motion describing the background fields on the target
space which is a trivial torus fibration over a $d$ dimensional space-time are given by the vanishing of the world-sheet beta-functions and can be
recovered, in the limit where we can ignore the $D$ internal coordinates, from a space-time Lagrangian $\mathscr{L}_{d}$. The supergravity has gauge group $U(1)^{2D}\subset O(D,D)$ and may be written in a
manifestly $O(D,D)$ invariant way
$$
\mathscr{L}_{d}=e^{-\phi}\left(R*1+*\d\phi\wedge \d\phi-\frac{1}{2}G_{(3)}\wedge *G_{(3)}+\frac{1}{2}\d{\cal M}^{IJ}\wedge *\d{\cal M}_{IJ}
 -\frac{1}{2}{\cal M}_{IJ}{\cal F}^I\wedge*{\cal F}^J\right)
$$
The scalars ${\cal M}_{IJ}$ take values in the coset space $O(D,D)/O(D)\times O(D)$. The details of this reduction are given in \cite{Maharana
``Noncompact symmetries in string theory'',Kaloper ``The O(dd) story of massive supergravity'',Hull ``Flux compactifications of string theory on
twisted tori''} and the conventions of \cite{Hull ``Flux compactifications of string theory on twisted tori''} have been used. This supergravity
may also be found directly by compactification of the ten (or twenty-six) dimensional supergravity on $T^{D}$ and truncating to the zero modes of
the harmonic expansions in the fields. Many examples of massive deformations, principally gaugings which preserve the maximal supersymmetry,
have been studied. For example see \cite{Hull ``Flux compactifications of string theory on twisted tori'',Hull ``Flux compactifications of
M-theory on twisted tori'',d=4,d=5,d=6,d=7,Bastian,RRE2b} and references therein. The examples of interest here are those in which a $2D$-dimensional group
$\cG$ is gauged\footnote{Note that $\cG$ is not a subgroup of $O(D,D)$. This is simply seen from the example $\cG=U(1)^{2D}$ discussed above. The Cartan torus of $O(D,D)$ is not large enough to contain the torus $T^{2D}$}. The resulting gauged theory is of the general form
\begin{eqnarray}\label{O(D-1,D-1) Lagrangian}
\mathscr{L}_{d}&=&e^{-\phi}\left(R*1+*\d\phi\wedge \d\phi+\frac{1}{2}{\cal H}_{(3)}\wedge *{\cal H}_{(3)}+\frac{1}{2}*D{\cal M}_{MN}\wedge D{\cal
M}^{MN}\right. \nonumber\\ &&\left.-\frac{1}{2}{\cal M}_{MN}{\cal F}^M\wedge*{\cal F}^N\right)+V*1
\end{eqnarray}
where the structure constants of the gauge algebra $[{\cal Z}_M,{\cal Z}_N]=t_{MN}{}^P{\cal Z}_P$ appear as massive deformation
parameters. For example, the scalar potential
$$
V=e^{-\varphi}\left(\frac{1}{4}{\cal M}^{MQ}L^{NT}L^{PS}t_{MNP}t_{QTS}- \frac{1}{12}{\cal M}^{MQ}{\cal M}^{NT}{\cal M}^{PS}t_{MNP}t_{QTS}\right)
$$
depends explicitly on the structure constants. The question of how such gauged supergravities can be realised in string theory and in particular
how to lift these supergravities to compactifications of ten and twenty-six dimensional theories has been the subject of much recent activity
\cite{Hull ``Flux compactifications of string theory on twisted tori'',Hull ``Flux compactifications of M-theory on twisted
tori'',Dall'Agata:2007sr}.

It has been suggested that many gauged supergravities cannot be realised as compactifications on manifolds in the
conventional sense but instead may be realised as `non-geometric compactifications', in which the $D$ dimensional compact space is not a
manifold, but a more general string background in which the duality symmetries of the theory play an important role \cite{Hull ``Flux compactifications of M-theory on twisted tori'',Dall'Agata:2007sr,Hull ``A geometry for
non-geometric string backgrounds'',Dab,Dabholkar ``Generalised T-duality and
non-geometric backgrounds''}. The prototypical example of such a non-geometric background is the T-fold \cite{Hull ``A
geometry for non-geometric string backgrounds''}. This is a $D$ dimensional background constructed as a $T^{D-1}$ fibred over a base circle. The
monodromy of the fibration is such that theory in the toroidal fibres is glued together, upon circumnavigating the base, by a transition
function which involves a T-duality (a general action of $O(D-1,D-1;\Z)$). The background is locally geometric but globally non-geometric.

The coordinate on the
base circle is taken to be $x\sim x+1$. Backgrounds of this form with monodromies in $O(D-1,D-1;\Z)$ give rise to gauged supergravities with gauge algebras
$$
[Z_x,Z_a]=f_{xa}{}^bZ_b+H_{xab}X^b   \qquad  [Z_x,X^a]=-f^a{}_{xb}X^b+c_x{}^{ab}Z_b
$$
\begin{equation}\label{fibration algebra}
[Z_a,Z_b ]=K_{xab}X^x  \qquad [X^a,Z_b]=-f^a{}_{xb}X^x \qquad  [X^a,X^b]=c_x{}^{ab}X^x
\end{equation}
with all other commutators vanishing and $a,b,=1,2,..D-1$. The generators $Z_x$ and $Z_a$ can be thought of as related to isometry generators
along the base circle and torus fibre respectively, whilst the generators $X^x$ (or $X^a$) are related to transformations of the $B$-field
components with one leg along the circle and the other in the $D-1$ dimensional torus. The structure constants characterize the
monodromy of the fibration. In particular $f_{xa}{}^b$ relates to an $SL(D-1;\Z)$ large dffeomorphisms arising from the topology of
the background, $H_{xab}$ is due to a monodromy in the $B$-field\footnote{The $B$-field is not globally defined, indicating a non-trivial field
strength $dB=H_{xab}dx\wedge dz^a\wedge dz^b+...$.} and $c_x{}^{ab}$ gives rise to a monodromy which includes a T-duality and indicates that the
internal space is a T-fold. More details on the origin and interpretation of this gauge algebra may be found in \cite{Hull:2007jy}. The action
of $O(D-1,D-1;\Z)$ along the fibre directions exchanges $Z_a$ and $X^a$ and gives different T-dual descriptions of the physics.
Since the fibres are tori the question of whether the action of $O(D-1,D-1;\Z)$ really is a duality depends on global issues and were studied in
\cite{Hull ``Global Aspects of T-Duality Gauged Sigma Models and T-Folds'',HullRRE}.

It was conjectured in \cite{Dabholkar ``Generalised T-duality and non-geometric backgrounds''} that the full action of $O(D,D;\Z)$, including
dualisation along the non-isometric $x$ (base coordinate) direction, may generate a genuine symmetry of the string theory. Dualising along the $x$ direction, exchanging $Z_x$ and $X^x$, produces a
supergravity with gauge algebra
$$
[X^x,Z_a]=c_{a}{}^{xb}Z_b+f_{ab}{}^xX^b   \qquad  [X^x,X^a]=-c^{xa}{}_{b}X^b+R^{xab}Z_b
$$
$$
[Z_a,Z_b ]=f_{ab}{}^xZ_x  \qquad [X^a,Z_b]=-c^{xa}{}_{b}Z_x \qquad  [X^a,X^b]=R^{xab}Z_x
$$
The background then includes the so-called $R$-flux, which refers to the structure constant $R^{xab}$. There is some evidence that
backgrounds with $R$-flux are not even locally geometric, although the precise nature of these backgrounds is yet to be understood. One may
think of such backgrounds as a $T^{D-1}$ fibration over the dual coordinate ${\ti x}$, conjugate to the winding modes along the base
\cite{Dabholkar ``Generalised T-duality and non-geometric backgrounds''}.

More generally, a supergravity of the form (\ref{O(D-1,D-1) Lagrangian}) that does not arise from a compactification on an internal space which is
a torus fibration admits gauge algebras that take the more general form
\begin{equation}\label{****}
[Z_m,Z_n]=f_{mn}{}^pZ_p+H_{mnp}X^p \qquad [Z_m,X^n]=\gamma^{np}{}_mZ_p+h_{mp}{}^nX^p \qquad [X^m,X^n]=c^{mn}{}_pX^p+R^{mnp}Z_p
\end{equation}
Note that in those cases where $H_{mnp}=R^{mnp}=0$, $\gamma_m{}^{np}=-c_m{}^{np}$, and $h_{mn}{}^p=f_{mn}{}^p$ the above gauge algebras are all
Drinfel'd doubles (\ref{**}) and arise from compactifications on backgrounds which satisfy the Poisson-Lie condition (\ref{poisson lie condition}). It is this class of gaugings we shall be prinicipally concerned with in this paper.

\subsection{Doubled Geometry}

In \cite{Hull ``A geometry for non-geometric string backgrounds''} a sigma model was proposed in which locally geometric $T^{D-1}$ fibrations
over a circle, with coordinate $x\sim x+1$, could be partially understood geometrically. In this approach the $T^{D-1}$ coordinates\footnote{For
a $T^{D-1}$ background the $z^a$ would be conjugate to momentum modes and the ${\ti z}_a$ would be conjugate to winding modes. However, it is not
clear what the analogue of momentum and winding modes are for the more general examples considered later.} $z^a$ are packaged together with the
coordinates ${\ti z}_a$ on the dual torus into $\mathbb{X}^I=(z^a,{\ti z}_a)$, where $I$ runs from $1$ to $2D-2$. The sigma model of \cite{Hull ``A geometry for non-geometric string backgrounds''} then
describes a world-sheet embedding into a $2D-1$ dimensional background with coordinates $(x,\mathbb{X}^I)$ where $x$ is again the base circle
coordinate and the $\mathbb{X}^I$ are coordinates on the doubled torus fibre. The doubled torus fibre has metric $\mathcal{M}_{IJ}$ given in
(\ref{M}) and generally depends on $x$. The correct number of physical degrees of freedom are ensured by the imposition of the constraint
\begin{equation}\label{constraint for doubled torus}
d\mathbb{X}^I=L^{IJ}\mathcal{M}_{JK}(x)*d\mathbb{X}^K+...
\end{equation}
where $L_{IJ}$ is the invariant of $O(D-1,D-1)$ and the ellipsis denotes terms involving other target space coordinates. The details of this construction are given in \cite{Hull ``A geometry for non-geometric string
backgrounds''}. The doubled formalism has proved to be a very useful tool in elucidating the structure of non-geometric backgrounds and
T-duality for various locally geometric torus fibrations however, as demonstrated in \cite{Hull:2007jy}, it has many limitations. In particular,
the doubled torus formalism gives a geometric interpretation for the monodromy of the fibration - as a large diffeomorphism of the doubled
geometry - but does not give a geometric interpretation of the gauge algebra (\ref{fibration algebra}). It is also unsuitable for various
backgrounds of interest which are not torus fibrations, in particular there is no way of incorporating any but the simplest Poisson-Lie
backgrounds into the doubled torus formalism as the general Poisson-Lie background is
not a torus fibration. In \cite{HullRRE} a sigma model in which all embedding coordinates are doubled was introduced so that $(x,z^a)\rightarrow \mathbb{X}^I=(x,{\ti x},z^a,{\ti z}_a)$. As we shall discuss in section three, this model and a chiral version of it describes, not only all backgrounds which satisfy (\ref{poisson lie condition}), but also generalisations which include $H$- and $R$-fluxes. The study of such backgrounds and their asociated sigma models is the subject of this paper.

In the next section we discuss certain geometric and algebraic structures relating to backgrounds that satisfy the conditions (\ref{poisson lie condition}) and the associated doubled geometries. Particular emphasis
is placed on the relationship between bi-algebras on $T\oplus T^*$ and Lie algebras on the doubled space. Section three begins by reviewing the
doubled worldsheet formalism introduced by Klim\v{c}ik and Severa in \cite{Klimcik:1995ux} and it is shown that this formalism is a
generalisation of a doubled formalism introduced by Tseytlin in \cite{Tseytlin:1990va}. Similar observations have been made in \cite{Dall'Agata:2008qz}. We also consider the sigma
model introduced in \cite{HullRRE} on the doubled space which generalises the sigma model of \cite{Hull ``A geometry for non-geometric string backgrounds''} and overcomes
the deficiencies of the sigma model of \cite{Klimcik:1995ux} by allowing the inclusion of $H$- and $R$-fluxes. Further details of some of the calculations are to be found in the Appendices.

One of the aims of this paper is to present a clear example of a string theory in which the target space can be explicitly treated from both the perspecive of generalised geometry and the doubled formalism reviewed above. Generalised geometry is a powerful tool for studying string backgrounds that include Riemannian target spaces as it is usually less difficult to construct the generalised bundle $T\oplus T^*$ than identify the correct doubled geometry. By contrast, the doubled geometry associated to backgrounds is unknown except for the very simplest cases. The power of the doubled formalism lies in its ability to explicitly describe non-geometric backgrounds, whereas we can currently only make somewhat vague assertions about non-geometric backgrounds from the perspective of structures on $T\oplus T^*$. It is hoped that an anlysis of examples where we can understand the background in both descriptions, such as presented here, will allow both perspectives to be generalised with confidence. Some hints at what such a generalisation may entail are discussed in the conclusion.

\section{Generalised geometry and doubled geometry}

The aim of this section is to provide a concise review of the algebraic and geometric structures that define the Poisson-Lie target spaces
discussed in later sections which will clarify the relationship of the sigma model in \cite{Klimcik:1995dy} with the doubled formalism of \cite{HullRRE}. The review is divided into three main parts. The staring point is the group manifold $G$ and the study of algebraic
structures which can be defined on the generalised tangent bundle $TG\oplus T^*G$. It will be shown how a compatible Poisson structure on $G$
allows an algebraic structure, analogous to the Lie bracket on $TG$, to be defined on $T^*G$. In the second part, an equivalent description of
the algebraic structures on $TG\oplus T^*G$ in terms of a doubled Lie algebra structure on doubled group $\cG=G\bowtie
\widetilde{G}$, where the Lie bracket on $T\widetilde{G}$ is related to the Poisson bracket on $G$, is considered. The spirit of these two
complementary approaches is indicative of the approaches to non-geometric backgrounds in string theory where one may use the $T\oplus T^*$
approach of \cite{Grana:2005ny,Grana:2006hr,Gurrieri:2002wz} or alternatively the doubled geometry approach of \cite{Dall'Agata:2007sr,Hull ``A geometry for
non-geometric string backgrounds'',Hull:2007jy}. The connection between these approaches is only understood in certain limited
examples and it is hoped that this section will at least elucidate the relationship between them for Poisson-Lie target spaces. The
third part of this section will consider explicit structures on the doubled geometry and the action of the group $O(D,D)$. This last section
will allow contact to be made with the doubled formalisms of \cite{Tseytlin:1990va} and \cite{Hull ``A geometry for non-geometric string
backgrounds''} in the next section, where sigma models on these backgrounds are considered. Appendix A contains an explicit example which
demonstrates some of the issues discussed in this section.

The focus of the discussion will be on those structures that play a significant role in string theory and many details of the constructions and
definitions will be glossed over or omitted entirely. In particular the discussion will be tailored to suit the considerations of the sigma
models in later sections. The interested reader may find further details in the general references \cite{Chari:1994pz,Majid:1996kd} and also the
original papers \cite{Drinfeld:1983ky,SemenovTianShansky:1993ws,Drinfeld:1986in,Kosmann-Schwarzback:1996}.

\subsection{The geometry of Lie groups}

The starting point is a $D$ dimensional Lie group $G$ with group multiplication $m:G\times G\rightarrow G$. The Lie algebra $\mathfrak{g}$
of $G$ has a natural bracket, the Lie bracket $[\,,\,]:\mathfrak{g}\times \mathfrak{g}\rightarrow \mathfrak{g}$ which encodes the local
structure of $G$. Let $\{T_m\}$ be a basis for a $D\times D$
matrix representation of $\mathfrak{g}$, and the Lie algebra may be written
\begin{equation}\label{*}
[T_m,T_n]=f_{mn}{}^pT_p
\end{equation}
where the associativity of the Lie group requires that the structure constants satisfy the Jacobi identity $f_{[mn}{}^qf_{p]q}{}^t=0$.

There is a natural action of $G_L\times G_R$ of the group acting from the left ($G_L$) and the right ($G_R$), on the group manifold $G$ given
by $G_L:g\rightarrow hg$ on the left and $G_R:g\rightarrow gh$ on the right, where $h$ is a constant element of G. For a bi-invariant metric,
such as the Cartan-Killing metric $\eta_{mn}=\frac{1}{2}f_{mp}{}^qf_{nq}{}^p$, the action of $G_L\times G_R$ is isometric. At any point $g\in G$
we can define left and right-invariant one-forms $\ell=g^{-1}\d g$ and $r=\d gg^{-1}$ respectively. These take values in $\mathfrak{g}\otimes T^*G$ and may be written as $\ell=\ell^mT_m$ and $r=r^mT_m$ where $\ell^m=\ell^m{}_i(x)\d x^i$ and $r^m=r^m{}_i(x)\d x^i$. The $x^i$ are local
coordinates on $G$, the index $i$ running from $1$ to $D$ and these one forms are globally defined on $G$. The algebraic structure of $G$ is encoded in
the Maurer-Cartan structure equations
\begin{equation}\label{structure equations}
\d\ell^m+\frac{1}{2}f_{np}{}^m\ell^n\wedge \ell^p=0  \qquad  \d r^m-\frac{1}{2}f_{np}{}^mr^n\wedge r^p=0
\end{equation}
It is in this way that the forms $\ell^m$ and $r^m$ encode information about the local structure of $G$. The natural pairing between vectors and
one-forms defines an invariant inner product $(\,|\,\,):TG\times T^*G\rightarrow\R$ such that $(dx^i|\partial_j)=\delta^i{}_j$. Left- and
right-invariant vector fields $K_m$ and $\widetilde{K}_m$, dual to the one-forms $\ell^m$ and $r^m$ respectively, can be defined which satisfy
$$
(\ell^m|K_n)=\delta^m{}_n   \qquad  (r^m|\widetilde{K}_n)=\delta^m{}_n
$$
from which it is easy to show
$$
K_m=(\ell^{-1})_m{}^i\frac{\partial}{\partial x^i} \qquad  \widetilde{K}_m=(r^{-1})_m{}^i\frac{\partial}{\partial x^i}
$$
The vector fields $K_m$ and $\widetilde{K}_m$ are invariant under the rigid action of $G_L$ and
$G_R$ respectively\footnote{Note that $K_m$ ($\widetilde{K}_m$) is invariant under the rigid action of $G_L$ ($G_R$) and generates the
infinitesimal action of $G_R$ ($G_L$), i.e. the left (right)-invariant generators are generators of the right (left) action.} and are globally
defined on $G$. Both the left and the right-invariant vector fields $K_m,\widetilde{K}_m\in TG$ each satisfy the Lie algebra
commutation relations (\ref{*}). The full isometry algebra $G_L\times G_R$, of the group manifold $G$ is generated by the vector fields $K_m$ and
$\widetilde{K}_m$
\begin{equation}\label{nilfold algebra}
[K_m,K_n]=f_{mn}{}^pK_p   \qquad  [K_m,\widetilde{K}_n]=0  \qquad  [\widetilde{K}_m,\widetilde{K}_n]=-f_{mn}{}^p\widetilde{K}_p
\end{equation}

\subsubsection{Example: Torus bundle}

As an example consider the group generated by
$$
[T_x,T_a]=-N_a{}^bT_b   \qquad  [T_a,T_b]=0
$$
where $a,b=1,2,..D-1$. This algebra can be represented by the $D\times D$ matrices
$$
T_x=\left(\begin{array}{cc}-N^a{}_b & 0 \\ 0 & 0
\end{array}\right)  \qquad  T_a=\left(\begin{array}{cc}0 & e_a \\ 0 & 0
\end{array}\right)
$$
where $e_a$ is the $D-1$-dimensional column vector with a 1 in the a'th position and zeros everywhere else.

This group is non-compact but can be compactified by identifying by an action of a discrete subgroup \cite{Hull ``Flux compactifications of
string theory on twisted tori''}. The manifold is then of the form $G/\G_{G}$ where the discrete subgroup $\G_{G}\subset G_L$ acts
from the left and is chosen such that $G/\G_{G}$ is compact. Such a $\G_{G}$, where it exists, is called cocompact and the space
$G/\G_{G}$ is generally not a group manifold. $G/\G_{G}$ is often, somewhat misleadingly, called a twisted torus, an example
of which is the dimensional nilmanifold. As a manifold, $G/\G_{G}$ can be constructed as a $T^{D-1}$ bundle over a circle with a monodromy
in the mapping class group\footnote{The discrete sub-group of diffeomorphisms not connected to the identity.} of $T^{D-1}$. One could equivalently
consider actions of a cocompact group from the right to define a manifold $G/\widetilde{\G}_{G}$ where $\widetilde{\G}_{G}\subset G_R$. The
convention here will be to consider the cocompact group to act from the left. Such global issues were investigated in \cite{HullRRE}.

This group can be viewed as a $T^{D-1}$ fibred over $S^1$ with monodromy given by $N^a_b$.  Coordinates $x^i=(x, z^a)$ can be introduced locally for the group manifold $G$, where $z^a$ are coordinates on the $T^{D-1}$ fibre and $x$ is the
coordinate along the base circle. A general element of the group
 $g\in G$
may then be given by
 \begin{equation}\label{groupG}
g=\left(\begin{array}{cc} \left(e^{-Nx}\right)^a{}_b & z^a \\ 0 & 1
\end{array}\right)
\end{equation}
and the left-invariant Maurer-Cartan forms, $\ell=g^{-1}\d g$  are given by
\begin{equation} \label{L forms}
\ell^x=\d x \qquad \ell^a=\left(e^{-Nx}\right)^a{}_b\d z^b
\end{equation}
The $\ell^m$ are dual to the left-invariant vector fields
$$
K_x=\left(e^{Nx}\right)_a{}^b\partial_b   \qquad K_a=\partial_a
$$
which generate the left-invariant part of the gauge algebra (\ref{nilfold algebra}). Right-invariant one-forms $r=\d gg^{-1}$ may also be
defined
\begin{equation} \label{R forms}
r^x=\d x \qquad r^a=\d z^a+N^a{}_bz^b\d x
\end{equation}
which are dual to the right-invariant vector fields
$$
\widetilde{K}_x=\partial_x-N_a{}^bz^a\partial_b   \qquad \widetilde{K}_a=\partial_a
$$
The global structure of $G/\G_G$ is given by the identifications of the coordinates
$$
x\sim x+\alpha  \qquad  z^a\sim (e^{N\alpha})^a{}_bz^b+\beta^a
$$
where $\alpha$ and $\beta^a$ are constants which specify $\G_G$. Since $\G_G$ acts
from the left, the left-invariant one-forms $\ell^m$ and vectors $K_m$ are globally defined on $G/\G_{G}$ whereas the right-invariant
$r^m$ and $\widetilde{K}_m$ generally will not be.

\subsection{Bi-algebra structures}

The Lie bracket $[\,,\,]:TG\times TG\rightarrow G$ and exterior derivative $\d:T^*G\rightarrow T^*G\times T^*G$ allow one to encode the algebraic
structure of $\mathfrak{g}$ in terms of either vectors (\ref{nilfold algebra}) or one-forms (\ref{structure equations}); however, there is a
clear asymmetry and one may wonder if operations $[\,,\,]^*:T^*G\times T^*G\rightarrow T^*G$ and $\delta: TG\rightarrow TG\times TG$ can be
defined to make the action of the set of maps more symmetric. This can be easily achieved if the group manifold has, in addition, a
Poisson structure that is compatible with the group action. Such groups are often called Poisson-Lie groups \cite{Chari:1994pz,Majid:1996kd}. A Poisson structure defines a bilinear map
$\{\,,\,\}:C^{\infty}(G)\times C^{\infty}(G)\rightarrow C^{\infty}(G)$ - the Poisson bracket - which may be written as
\begin{equation}\label{poisson bracket}
\{f,f'\}=\pi^{ij}\partial_if\partial_jf'
\end{equation}
where $f,f'\in C^{\infty}(G)$ are functions on the group manifold and $\pi^{ij}$ defines a Poisson bi-vector
$$
\pi=\frac{1}{2}\pi^{ij}\partial_i\wedge\partial_j
$$
so that (\ref{poisson bracket}) may then be written as $\{f,f'\}=(\pi|\d f\wedge \d f')$. A smooth map $\mathscr{F}:G_1\rightarrow G_2$ between
Poisson manifolds $G_1$ and $G_2$ is a Poisson map if it preserves the Poisson brackets on $G_1$ and $G_2$, i.e. if
$$
\{\mathscr{F}(f_1),\mathscr{F}(f'_1)\}_2=\mathscr{F}(\{f_1,f'_1\}_1)
$$
where $f_1,f'_1\in C^{\infty}(G_1)$ and $\mathscr{F}(f_1),\mathscr{F}(f'_1)\in C^{\infty}(G_2)$. $\{,\}_1$ and $\{,\}_2$ are Poisson brackets on
$G_1$ and $G_2$ respectively. A Poisson-Lie group is a Lie group with a compatible Poisson structure, i.e. the group multiplication $m:G\times
G\rightarrow G$ is a Poisson map.

\subsubsection{Lie bi-algebras}

Let $\d f(e),\d f'(e)\in T^*G|_e$ be one-forms evaluated at the identity ($e$) given by the functions $f,f'\in C^{\infty}(G)$. A Poisson structure on $G$
induces a bracket, at the identity, $[\,,\,]^*:T^*G\times T^*G\rightarrow T^*G$ as
$$
[\d f(e),\d f'(e)]^*=(\d\{f,f'\})(e)
$$
More generally, we can write $[\ell^m(e),\ell^n(e)]^*=d(\pi|\ell^m\wedge \ell^n)(e)$. Using $(\pi|\ell^m\wedge \ell^n)=\pi^{ij}\ell^m{}_i\ell^n{}_j$ and using that $\pi^{ij}(e)=0$ one can write
$$
[\ell^m(e),\ell^n(e)]^*=(\d\pi^{ij}\ell^m{}_i\ell^n{}_j)(e)
$$
If we then define
$$
\left((\ell^{-1})_p{}^k\partial_k\pi^{ij}\ell^n{}_i\ell^p{}_j\right)(e)=c_p{}^{mn}
$$
where $c_m{}^{np}$ is a constant as the expression is evaluated at the identity. At the identity $g=e$ the co-bracket $[\,,\,]^*$ may then be
defined as
$[\ell^m(e),\ell^n(e)]^*=c_p{}^{mn}\ell^p(e)$. The definition of this bracket at the identity may then be extended globally by the action of the group to give
$$
[\ell^m,\ell^n]^*=c_p{}^{mn}\ell^p
$$
The bracket $[\,,\,]^*$ can be associated, via the inner product $(\,|\,)$, to a derivation $\delta:TG\rightarrow TG\times TG$ as
$(K_m|[\ell^n,\ell^p]^*)=(\delta K_m|\ell^n\wedge\ell^p)=c_m{}^{np}$. It is not hard to show that
$$
\delta K_m+\frac{1}{2}c_m{}^{np}K_n\wedge K_p=0
$$
The set $(TG\oplus T^*G,[\,,\,], [\,,\,]^*)$ defines a bi-algebra arising from the Lie algebra structures $(TG,[\,,\,])$ and $(T^*G,[\,,\,]^*)$.
It follows that there is a natural bracket on $TG\oplus T^*G$. The adjoint action of $G$ on $TG$ is given by the Lie bracket $Ad_KK'={\cal L}_{K'}K=[K,K']$.
The dual of this adjoint action, with respect to the inner product $(\,,\,)$ is the co-adjoint action $Ad^*$ of $G$ on $T^*G$, where
$$
(\ell|Ad_KK')=-(Ad^*_K\ell|K')
$$
Using the basis introduced above, this may be written as
$$
Ad_{K_m}K_n=[K_m,K_n]=f_{mn}{}^pK_p \qquad  Ad^*_{K_m}\ell^n={\cal L}_{K_m}\ell^n=-f_{mp}{}^n\ell^p
$$
Similarly one can define \cite{Kosmann-Schwarzback:1996} $Ad_{\ell}\ell'=[\ell,\ell']^*$ and the dual operation with respect to the inner
product
$$
(Ad_{\ell}\ell'|K)=-(\ell'|Ad^*_{\ell}K)
$$
Let $[\,,\,]_{\mathfrak{d}}$ denote the bracket on $TG\oplus T^*G$ where $[K,K']_{\mathfrak{d}}=[K,K']$ and
$[\ell,\ell']_{\mathfrak{d}}=[\ell,\ell']^*$ then, using the fact that the inner product $(\,|\,)$ is invariant under the adjoint action, it is
not hard to show that $([\ell,K]_{\mathfrak{d}}|K')=-(Ad^*_K\ell|K')$ and $([\ell,K]_{\mathfrak{d}}|\ell')=(Ad^*_{\ell}K|\ell')$ so that the
cross term for the bracket is
$$
[\ell,K]_{\mathfrak{d}}=Ad^*_{\ell}K-Ad^*_K\ell
$$
In terms of the basis introduced here
$$
Ad_{\ell^m}\ell^n=[\ell^m,\ell^n]^*=c_p{}^{mn}\ell^p    \qquad  Ad_{\ell^m}K_n={\cal L}^*_{\ell^m}K_n=-c_n{}^{mp}K_p
$$
From now on, the bracket $[\,,\,]_{\mathfrak{d}}$ will simply be referred to as $[\,,\,]$.

For $\pi^{ij}$ a constant, and so $c_m{}^{np}=0$, the appropriate bracket on $TG\oplus T^*G$ is the Courant bracket \cite{Courant}. For $c_m{}^{np}\neq 0$ the natural
bracket is a generalization of the Courant bracket given by
\begin{equation}\label{a}
[K+\ell,K'+\ell']=[K,K']+[\ell,\ell']^*-Ad^*_K\ell'+Ad^*_{K'}\ell+Ad^*_{\ell'}K-Ad^*_{\ell}K'
\end{equation}
where $Ad^*_K$ is the usual adjoint action of $G$ on $TG$ and $Ad^*_{\ell}$ is the co-adjoint action of $G$ on $T^*G$.

Recall that the canonical pairing between forms $\ell^m$ and vectors $K_m$ on $TG\oplus T^*G$ defines a natural inner product $(\,|\,\,)$ on
$\mathfrak{g}\oplus\mathfrak{g}^*$. For two generalized vectors $Y=\ell+K$ and $Y'=\ell'+K'$ in $TG\oplus T^*G$ the
inner product is
$$
(Y|Y')=(\ell|K')+(\ell'|K)
$$
This defines an $O(D,D)$ structure on $TG\oplus T^*G$ which preserves the inner product.

\subsection{Drinfel'd Doubles, Manin Triples and Doubled Geometry}

The definition of a Lie bi-algebra given above hides the symmetric role that $TG$ and $T^*G$ play. A more symmetric description is given in
terms of Drinfel'd doubles and Manin triples. A Drinfel'd double is a Lie algebra $\mathfrak{h}$ with inner product such that
$$
\mathfrak{h}=\mathfrak{g}\oplus\widetilde{\mathfrak{g}}
$$
where the sub-algebras $\mathfrak{g}$ and $\widetilde{\mathfrak{g}}$ are maximally isotropic with respect to the inner product. The Drinfel'd
double $\mathfrak{h}$ can be integrated up to give a Lie-group $\cG=G\bowtie\widetilde{G}$, for which the vectors of the tangent space
satisfy the commutation relations of the Drinfel'd double $T\cG$ (\ref{**}) \cite{Drinfeld:1986in,Kosmann-Schwarzback:1996}. Consider
the Lie algebra $T\widetilde{G}$ associated with $T^*G$ where the Lie bracket on $T\widetilde{G}$ is associated with the dual bracket
$[\,,\,]^*$ described above. Let $X^m$ be a $\cG_L$-invariant basis for $T\widetilde{G}$ so that the dual algebra may be written as
$$
[X^m,X^n]=c^{mn}{}_pX^p
$$
and there is a similar commutator for the right-invariant basis with elements $\widetilde{X}^m$. The left-invariant ${\cal Z}_M$ and
right-invariant $\widetilde{\cal Z}_M$ vectors may then be written as
\begin{eqnarray}\label{T}
{\cal Z}_M=\left(%
\begin{array}{c}
  Z_m \\
  X^m \\
\end{array}%
\right) \qquad  \widetilde{\cal Z}_M=\left(%
\begin{array}{c}
  \widetilde{Z}_m \\
  \widetilde{X}^m \\
\end{array}%
\right)
\end{eqnarray}
where $Z_m$ and $\widetilde{Z}_m$ are the lifts of the left and right-invariant isometry generators $K_m$ and $\widetilde{K}_m$ in $TG$ to
$T\cG$. The full Lie-algebra of $\cG$ is then $[{\cal Z}_M,{\cal Z}_N]=t_{MN}{}^P{\cal Z}_P$, or in terms of the left-invariant
generators of the maximally isotropic sub-algebras $G_L$ and $\widetilde{G}_L$
$$
[Z_m,Z_n]=f_{mn}{}^pZ_p \qquad [Z_m,X^n]=c^{np}{}_mZ_p-f_{mp}{}^nX^p \qquad [X^m,X^n]=c^{mn}{}_pX^p
$$
There is a similar set of commutators for the right-invariant generators $(\widetilde{Z}_m,\widetilde{X}^m)$ of $G_R$ and $\widetilde{G}_R$.
Associativity of $\cG$ requires that the structure constants $t_{MN}{}^P$ satisfy\footnote{The convention for symmetrization and
anti-symmetrization of indices is with weight one, for example: $A^{[mn]}=\frac{1}{2}\left(A^{mn}-A^{nm}\right)$ and
$A^{(mn)}=\frac{1}{2}\left(A^{mn}+A^{nm}\right)$.}
$$
f_{[mn}{}^qf_{p]q}{}^t=0    \qquad c^{[mn}{}_qc^{p]q}{}_t=0   \qquad  f_{qp}{}^mc^{rs}{}_m=4c^{m[s}{}_{[p}f_{q]m}{}^{r]}
$$
Left-invariant one-forms ${\cal P}^M$, dual to the left-invariant vectors ${\cal Z}_M$, can be defined in terms of the elements $h\in \cG$ of
the doubled group where
$$
(\mathcal{P}^M|{\cal Z}_N)=\delta^M{}_N    \qquad  \mathcal{P}^M=(h^{-1}\d h)^M
$$
Similarly, one can define right-invariant vectors $\widetilde{\mathcal{P}}^M=(\d hh^{-1})^M$ dual $\widetilde{\cal Z}_M$. In terms of the local
coordinates $\mathbb{X}^I$ on $\cG$, where $I$ runs from $1$ to $2D$, the one-forms can be written as
\begin{equation}\label{doubled forms}
\mathcal{P}^M=\mathcal{P}^M{}_I\d\mathbb{X}^I    \qquad  \widetilde{\mathcal{P}}^M=\widetilde{\mathcal{P}}^M{}_I\d\mathbb{X}^I
\end{equation}
and the corresponding vector fields are
$$
{\cal Z}_M=(\mathcal{P}^{-1})_M{}^I\partial_I    \qquad  \widetilde{\cal Z}_M=(\widetilde{\mathcal{P}}^{-1})_M{}^I\partial_I
$$
where $\partial_I$ denotes partial differentiation with respect to $\mathbb{X}^I$.

The bi-algebra describes structures on $T\oplus T^*$ and is a specific case of the more general field of generalised geometry
\cite{Gualtieri:2003dx,Gualtieri:2007ng}. By contrast, the Drinfel'd double replaces $T\oplus T^*$ with the Lie algebra structures on a doubled group
$\cG$ and falls into the general scheme of the doubled formalism approach, often used in the discussion of non-geometric backgrounds in
string theory \cite{Hull ``A geometry for non-geometric string backgrounds'',Hull:2007jy}. The connection
between the two approaches is simple in this case and is fixed by the constraint $({\cal P}^M|{\cal P}^N)=L^{MN}$. This condition allows forms
in $T^*\widetilde{G}$ to be identified with vectors in $TG$ and so the $2D$ forms on the double $T^*\cG$ may be re-expressed in terms of
generalised vectors on $TG\oplus T^*G$ (see \cite{Dall'Agata:2007sr} for some worked examples). A simple example is $\cG=\R^{2D}$,
then $(\d\mathbb{X}^I|\d\mathbb{X}^J)=L^{IJ}$ gives $(\d x^i|\d\tilde{x}_j)=\delta^i{}_j$ therefore $\d\tilde{x}_i$ can be identified with
$\partial_i$ and the description in terms of $T^*\cG$ is replaced with an equivalent description in terms of $TG\oplus T^*G$. The
constructions of doubled formalisms in more general backgrounds, which do not have such a Lie group structure, and the relationship with
generalised geometry in such cases has been discussed in \cite{Dall'Agata:2007sr}.

\subsection{Manin triples and polarizations}

A Manin triple is a set of Lie algebras $(\mathfrak{h},\mathfrak{g},\tilde{\mathfrak{g}})$ with an inner product such that
$$
\mathfrak{h}=\mathfrak{g}\oplus\tilde{\mathfrak{g}}
$$
and $\mathfrak{g}$ and $\tilde{\mathfrak{g}}$ are maximally isotropic sub-algebras of $\mathfrak{h}$ with respect to that inner product. Let $\{T_m\}$ ($\{\tilde{T^m}\}$) be a basis of matrix generators for the sub-algebra $\mathfrak{g}$ ($\widetilde{\mathfrak{g}}$) and $\langle\,|\,\,\rangle$ an inner product
such that
\begin{equation}\label{K norm}
\langle T_m|T_n\rangle=0  \qquad \langle\widetilde{T}^m|\widetilde{T}^n\rangle=0    \qquad  \langle T_m|\widetilde{T}^n\rangle=\delta_m{}^n
\end{equation}
The inner product is adjoint-invariant $\langle
g^{-1}Ag|B\rangle=\langle A|gBg^{-1}\rangle$ and defines an $O(D,D)$ structure on $\mathfrak{g}\oplus \tilde{\mathfrak{g}}$. In this
representation the Drinfel'd double may be written
$$
[T_m,T_n]=f_{mn}{}^pT_p \qquad [T_m,\widetilde{T}^n]=c^{np}{}_mT_p-f_{mp}{}^n\widetilde{T}^p \qquad
[\widetilde{T}^m,\widetilde{T}^n]=c^{mn}{}_p\widetilde{T}^p
$$
The Drinfel'd double $\mathfrak{h}=\mathfrak{g}\oplus\tilde{\mathfrak{g}}$ may be decomposed in many different ways into many different
Manin triples $(\mathfrak{h},\mathfrak{g}_1, \tilde{\mathfrak{g}}_1)=(\mathfrak{h},\mathfrak{g}_2, \tilde{\mathfrak{g}}_2)=...$. The
possibility of many inequivalent decompositions leads to the notion of plurality \cite{Von Unge:2002ss}, involving many different pairs
$(\mathfrak{g}, \tilde{\mathfrak{g}})$, described by the same Drinfel'd double. Given an algebra $\mathfrak{h}$, the choice of a Manin triple is equivalent to choosing a polarization of the Lie algebra in the sense of
\cite{Hull:2007jy}. Thus different Manin triples select different sub-spaces in $\cG$ which we identify with with physical space-time
$G$. An example is $\cG=\R^{2D}$ where the different Manin triples - different ways of embedding $T^D\subset T^{2D}$ - are related to
each other by the action of $O(D,D;\Z)$ which is identified as T-duality. For any given Manin triple
$(\mathfrak{h},\mathfrak{g},\tilde{\mathfrak{g}})$ there exists at least one other triple
$(\mathfrak{h},\tilde{\mathfrak{g}},\mathfrak{g})$ given by exchanging $\mathfrak{g}$ and $\tilde{\mathfrak{g}}$. This $\Z_2$ symmetry of $\mathfrak{h}$ is the source of the Poisson-Lie map.

The group $G$ is recovered from the double $\cG$ by the left quotient $G=\cG/\widetilde{G}_L$. The
quotient restricts to the subspace which is invariant under the action of $\widetilde{X}^m$ - the generators of $\widetilde{G}_L$. On the coset
$G=\cG/\widetilde{G}_L$ the group $\widetilde{G}_L$ has no action and the left-invariant generators $Z_m$ reduce to the isometry
generators $K_m$ of the group $G$. The coset is taken from the left and even though there is no action of $\widetilde{G}_L$ on $\cG/\widetilde{G}_L$, there may
still be an action of $\widetilde{G}_R$ on $\cG/\widetilde{G}_L$. In fact, the generators $X^m$ of $\widetilde{G}_R$ generate the isotropy group of the coset
$G=\cG/\widetilde{G}_L$. Alternatively, one could consider the Manin triple which exchanges $G$ with $\widetilde{G}$ and recover
$\widetilde{G}$ from $\cG$ as the left-acting coset $\widetilde{G}=\cG/G_L$. In the following
sections examples of complementary Manin triples $(\mathfrak{h},\mathfrak{g},\tilde{\mathfrak{g}})$ and
$(\mathfrak{h},\tilde{\mathfrak{g}},\mathfrak{g})$ will be considered and sigma models describing a world-sheet embedding into
$\cG/\widetilde{G}_L$ and $\cG/G_L$ constructed. The map between these sigma models on
$\cG/\widetilde{G}_L$ and $\cG/G_L$ is the Poisson-Lie map. This map is often, erroneously,
referred to as a T-duality but, as stated in the introduction, it is generally not a duality of the sigma model and so the term Poisson-Lie
\emph{map} will be used instead.

\subsection{Adjoint actions on the double}

Once a Manin triple (or polarization) $(\mathfrak{h},\mathfrak{g},\tilde{\mathfrak{g}})$ has been chosen, one may meaningfully distinguish between the
sets of generators $T_m$ and $\widetilde{T}^m$. The generators for the double, in the matrix
representation, are
$$
T_M=\left(%
\begin{array}{c}
  T_m \\
  \widetilde{T}^m \\
\end{array}%
\right)
$$
which generate the Lie-algebra of $\cG$
$$
[T_M,T_N]=t_{MN}{}^PT_P
$$
It is helpful to use a bra-ket notation in which a basis element of the Lie algebra is represented as $|T_M\rangle$, an operator on
the vector space as $|T_M\rangle\langle T_N|$ and so on. The inner product $\langle\, |\,\, \rangle$ to which the two Lie
sub-algebras, considered as vector spaces, are null may then be written as
$$
\langle T_M |T_N \rangle=L_{MN}
$$
where $L_{MN}$ is a metric on $\cG$ relating vectors $|\,\,\rangle$ and their duals $\langle\,|$ and is the invariant of $O(D,D)$
$$
L_{MN}=\left(%
\begin{array}{cc}
  0 & \bid_d \\
  \bid_d & 0 \\
\end{array}%
\right)
$$
An orthonormal basis may be defined in terms of the dual vector $\langle T^M|=\langle T_N|L^{MN}$ so that $\langle
T^M|T_N\rangle =\delta^M{}_N$.

The adjoint action of $G$ on the Lie algebra $T\cG$ is given by \footnote{The condition $\langle T_m|\widetilde{T}^n\rangle=\langle
g^{-1}T_mg|g^{-1}\widetilde{T}^ng\rangle=\delta_m{}^n$ has been used to simplify the expression.}
\begin{eqnarray}\label{adjoint action}
g^{-1}\left(%
\begin{array}{c}
  T_m \\
  \widetilde{T}^m \\
\end{array}%
\right)g=\left(%
\begin{array}{cc}
  A_m{}^n & 0 \\
  \beta^{mn} & (A^{-1})^m{}_n \\
\end{array}%
\right)\left(%
\begin{array}{c}
  T_n \\
  \widetilde{T}^n \\
\end{array}%
\right)
\end{eqnarray}
and similarly for the adjoint action of the $\widetilde{G}$ on $T\cG$
$$
\tilde{g}^{-1}\left(%
\begin{array}{c}
  T_m \\
  \widetilde{T}^m \\
\end{array}%
\right)\tilde{g}=\left(%
\begin{array}{cc}
  (\widetilde{A}^{-1})_m{}^n & \widetilde{\beta}_{mn} \\
  0 & \widetilde{A}^m{}_n \\
\end{array}%
\right)\left(%
\begin{array}{c}
  T_n \\
  \widetilde{T}^n \\
\end{array}%
\right)
$$
Requiring that $\beta^{mn}=-\beta^{nm}$, ensures that the adjoint matrix is an element of $O(D,D)$. This can be written as
$$
g^{-1}T_Mg={\cal O}_M{}^NT_N    \qquad  \tilde{g}^{-1}T_M\tilde{g}=\widetilde{{\cal O}}_M{}^NT_N
$$
where ${\cal O}$ and $\widetilde{{\cal O}}$ are elements of $O(D,D)$ and depend on $g$ and $\ti g$ respectively. Furthermore the condition
$\langle \widetilde{T}^m|\widetilde{T}^n\rangle=\langle
g^{-1}\widetilde{T}^mg|g^{-1}\widetilde{T}^ng\rangle=2(A^{-1})^{(m}{}_p\beta^{n)p}=0$ implies that the combination
$(A^{-1})^{m}{}_p\beta^{np}$ is anti-symmetric. In fact,one can simply define a pair of Poisson structures on $G$ and $\widetilde{G}$
\begin{equation}
\pi^{mn}(g)=(A^{-1})^{m}{}_p\beta^{np}  \qquad  \tilde{\pi}_{mn}({\ti g})=(A^{-1})_{m}{}^p\widetilde{\beta}_{np}
\end{equation}
These Poisson structures are compatible with the Lie group structures on $G$ and $\widetilde{G}$ respectively. Note that, at the identity
$\beta^{mn}(e)=\widetilde{\beta}_{mn}(e)=0$, so that $\pi^{-1}$ and $\tilde{\pi}^{-1}$ do not exist globally and so the Poisson-Lie manifolds are
not symplectic \cite{Chari:1994pz,Kosmann-Schwarzback:1996}. It is not too hard to show that the derivative of the Poisson structure, evaluated
at the identity, gives the structure constants of the double. For example, using the fact that $\beta^{mn}(e)=0$, it is simple to show that
$\partial_i\pi^{mn}|_e=\partial_i\beta^{mn}|_e$. From the above definition $\beta^{mn}=\langle g^{-1}\widetilde{T}^mg|\widetilde{T}^n\rangle$
\begin{equation}
\partial_i\beta^{mn}=\ell^q{}_i(\beta^{mp}f_{pq}{}^n-(A^{-1})_p{}^mc_q{}^{pn})
\end{equation}
so that $(\ell^{-1})_p{}^i\partial_i\pi^{mn}|_e=(\ell^{-1})_p{}^i\partial_i\beta^{mn}|_e=-c_p{}^{mn}$ as anticipated. The adjoint action of $G$
maps the right-invariant $r$ onto the left-invariant $\ell$ since $g^{-1}rg=\ell$. In components this reads
$r^mg^{-1}T_mg=\ell^mT_m=r^mA_m{}^nT_n$, so that $r^n{}_iA_n{}^m=\ell^m{}_i$. The adjoint action of $G$ may be written
simply as
$$
A_m{}^n(g)=(r^{-1})^i{}_m\ell^n{}_i
$$
Similarly for the adjoint action of $\widetilde{G}$
$$
\widetilde{A}^m{}_n(\tilde{g})=(\tilde{r}^{-1})_i{}^m\ell_n{}^i
$$

\subsection{Metrics}

The inner product $L_{MN}$, with signature $(D,-D)$ reduces the structure group of the double to $O(D,D)\subset GL(2D)$. The structure group may
be further reduced to $O(D)\times O(D)\subset O(D,D)$ by choosing a $D$-dimensional sub-bundle in $TG \oplus T\widetilde{G}$ on which the inner
product is positive definite. Denoting such a choice of sub-bundle by $\mathscr{E}^+$ and its complement in $\mathfrak{g} \oplus \tilde{\mathfrak{g}}$
by $\mathscr{E}^-$, then there is a splitting
$$
\mathfrak{g} \oplus \tilde{\mathfrak{g}}=\mathscr{E}^+\oplus\mathscr{E}^-
$$
This splitting defines a linear map ${\cal R}:\mathfrak{g} \oplus \tilde{\mathfrak{g}}\rightarrow \mathfrak{g} \oplus \tilde{\mathfrak{g}}$. ${\cal R}$ is idempotent, i.e ${\cal R}^2=1$ and thus defines an almost product structure. The two eigenspaces $\mathscr{E}^+$
($\mathscr{E}^-$) are positive (negative) definite with respect to the inner product $L$ and are associated with the eigenvalues $\pm 1$ of
${\cal R}$. More importantly, the eigenspaces $\mathscr{E}^{\pm}$ are the graph of $g\pm B:\widetilde{T}\mapsto T$ and may be written as
$$
\mathscr{E}^+=Span\{T_m+E_{mn}\widetilde{T}^n\}  \qquad  \mathscr{E}^-=Span\{T_m-E^t_{mn}\widetilde{T}^n\}
$$
where $E_{mn}=g_{mn}+B_{mn}$ is the background tensor evaluated at the identity and is therefore independent of the local coordinates $\mathbb{X}^I$.
$E^t_{mn}=E_{nm}=g_{mn}-B_{mn}$ denotes the transpose. $Span\{...\}$ can thought of the minimal set of possible linear combinations of the
elements (the set is taken to be minimal so that it forms a basis). Let $|\mathscr{E}^{\pm}{}_m\rangle$ denote bases for the eigenspaces
$\mathscr{E}^{\pm}$ given by
$$
|\mathscr{E}^+{}_m\rangle=\frac{1}{\sqrt{2}}\left(|T_m\rangle+E_{mn}|\widetilde{T}^n\rangle\right)    \qquad
|\mathscr{E}^-{}_m\rangle=\frac{1}{\sqrt{2}}\left(|T_m\rangle-E_{nm}|\widetilde{T}^n\rangle\right)
$$
It is also useful to define
$$
|\mathscr{E}^{\pm m}\rangle=g^{mn}|\mathscr{E}^{\pm}{}_n\rangle
$$
Using the properties (\ref{K norm}) of the inner product on $\mathfrak{g}\oplus\widetilde{\mathfrak{g}}$, the normalisation of this basis is
$$
\langle \mathscr{E}^{\pm m}|\mathscr{E}^{\pm}{}_n{}\rangle=\pm\delta^m{}_n    \qquad  \langle \mathscr{E}^+{}_m|\mathscr{E}^-{}_n\rangle=0
$$
so that the eigenspaces are indeed pseudo-orthonormal - the inner product is positive (negative) definite on the subspaces $\mathscr{E}^+$
($\mathscr{E}^-$). The explicit form of the product structure ${\cal R}$ can be written in terms of this basis
$$
{\cal R}=|\mathscr{E}^{+m}\rangle \langle\mathscr{E}^+{}_m|+|\mathscr{E}^{-m}\rangle \langle\mathscr{E}^-{}_m|
$$
Note also that the $2D\times 2D$ identity matrix can be written in this basis as
$$
\bid=|\mathscr{E}^{+m}\rangle \langle\mathscr{E}^+{}_m|-|\mathscr{E}^{-m}\rangle \langle\mathscr{E}^-{}_m|
$$
The matrix representation of the linear and idempotent map ${\cal R}$ is given by
$$
{\cal R}^M{}_N=\langle T^M|{\cal R}|T_N\rangle
$$
and a metric ${\cal M}_{MN}=L_{MP}{\cal R}^P{}_N$ may be defined and written as
$$
{\cal M}_{MN}=\langle T_M|{\cal R}|T_N\rangle
$$
It is actually the metric ${\cal M}$, which takes values in the coset $O(D,D)/O(D)\times O(D)$, that plays a more fundamental role in the later
sections of the paper. The determination of its explicit form in terms of the background metric and $B$-field is given in Appendix C and the result is quoted here
\begin{eqnarray}\label{M matrix}
{\cal M}_{MN}=\left(%
\begin{array}{cc}
  g_{mn}+B_{mp}g^{pq}B_{qn} & g^{np}B_{pm} \\
  g^{mp}B_{pn} & g^{mn} \\
\end{array}%
\right)
\end{eqnarray}
The corresponding product structure is given by
\begin{eqnarray}
{\cal R}^M{}_N=\left(%
\begin{array}{cc}
  g^{mp}B_{pn} & g_{mn}+B_{mp}g^{pq}B_{qn} \\
  g^{mn} & g^{np}B_{pm} \\
\end{array}%
\right)
\end{eqnarray}
 The action of $O(D,D)$ on ${\cal M}$ has a natural expression in terms of the adjoint action of $G$ (see Appendix D for further details)
\begin{equation}\label{odd}
{\cal M}_{MN}(g)=\langle g^{-1}T_M g|{\cal R}|g^{-1}T_N g\rangle ={\cal O}_M{}^P(g){\cal M}_{PQ}(e){\cal O}^Q{}_N(g)
\end{equation}
Let ${\cal M}(E)$ be the value of the metric ${\cal M}$ evaluated with entries $g_{mn}=E_{(mn)}$ and $B_{mn}=E_{[mn]}$, then the natural action
of $O(D,D)$ on the background is ${\cal M}\rightarrow {\cal O}{\cal M}{\cal O}^T$, where ${\cal O}\in O(D,D)$ is comprised of the following
parts:

\textbf{$GL(D)$ Transformations}
$$
{\cal O}_A=\left(%
\begin{array}{cc}
  A & 0 \\
  0 & (A^{-1})^T \\
\end{array}%
\right) \qquad  g\rightarrow AgA^T  \quad   B\rightarrow ABA^T
$$
where $A\in GL(D)$. The metric on the double then transforms under ${\cal O}_A$ as
$$
{\cal O}_A{\cal M}(E){\cal O}_A^T={\cal M}(AEA^T)
$$

\textbf{$b$ Transformations}
$$
{\cal O}_b=\left(%
\begin{array}{cc}
  1 & -b \\
  0 & 1 \\
\end{array}%
\right) \qquad  g\rightarrow g  \quad   B\rightarrow B+b
$$
where $b\in T^*G\wedge T^*G$. The metric on the double then transforms under ${\cal O}_b$ as
$$
{\cal O}_b{\cal M}(E){\cal O}_b^T={\cal M}(E+b)
$$

\textbf{$\beta$ Transformations}

$$
{\cal O}_{\beta}=\left(%
\begin{array}{cc}
  1 & 0 \\
  -\beta & 1 \\
\end{array}%
\right) \qquad  E\rightarrow (E^{-1}+\beta)^{-1}
$$
where $\beta\in TG\wedge TG$. The metric on the double then transforms under ${\cal O}_{\beta}$ as
$$
{\cal O}_{\beta}{\cal M}(E){\cal O}_{\beta}^T={\cal M}((E^{-1}+\beta)^{-1})={\cal M}^{-1}(E^{-1}+\beta)
$$
This coincides with the action of $O(D,D)$ on $T\oplus T^*$ studied in \cite{Gualtieri:2003dx,Gualtieri:2007ng}.

\subsection{Twisted tori and the global structure of the double}

So far the group $\cG$ has only been defined in terms of its Lie algebra - the Drinfel'd double - and therefore only its local structure
has been specified. For the considerations in the following sections, the target space need only locally be a group and globally it may have the
form \cite{Hull ``Flux compactifications of string theory on twisted tori''}
$$
\cX=\cG/\G_{\cG}
$$
where $\G_\cG$ is a discrete sub-group of $\cG_L$ such that $\cX$ is compact where $\G_\cG$ acts from the left. As discussed above, such
spaces are often called twisted tori and this nomenclature, though misleading, has now become standard and will be adopted here. Both left and
right-invariant objects will be globally defined on $\cG$ but only the left-invariant objects will generally be globally defined on $\cX$.
For the most part we shall ignore such global issues as they are treated at length in \cite{HullRRE} and the analysis there can straightforwardly be applied to all cases considered here.

\subsection{H-twisted bi-algebras}

As discussed further in Appendix A, the natural (bi-algebra) bracket on $TG\oplus T^*G$ for the case in which the Poisson structure on $G$ is
trivial ($\pi=0$) is the Courant bracket \cite{Courant}. In this section it is shown that the $H$-twisted Courant bracket \cite{Gualtieri:2003dx,Gualtieri:2007ng} is equivalent to the
$H$-twisted Drinfel'd double $\mathfrak{h}_H$. A natural basis of left-invariant one-forms and vectors on $TG\oplus T^*G$ is given by
$K_m=(\ell^{-1})_m{}^i\partial_i\in TG$ and $\ell^m=\ell^m{}_i\d x^i\in T^*G$. Consider a general element of $TG\oplus T^*G$
$$
Y=\xi^mK_m+\eta_m\ell^m \in TG\oplus T^*G
$$
where
$$
\imath_m\ell^n=( K_m|\ell^n)=\delta_m{}^n \qquad (Y|\bar{Y})=\xi^m\bar{\eta}_m+\eta_m\bar{\xi}^m
$$
and $\xi^m$ and $\eta_m$ are some arbitrary constant parameters\footnote{Or, equivalently, general parameters evaluated at the identity.}. The
twisted Courant bracket \cite{Gualtieri:2003dx,Gualtieri:2007ng} is
$$
[Y,\bar{Y}]_H=[\xi,\bar{\xi}]+{\cal L}_{\xi}\bar{\eta}-{\cal
L}_{\bar{\xi}}\eta-\frac{1}{2}\d\left(\iota_{\xi}\bar{\eta}-\iota_{\bar{\xi}}\eta\right)+\imath_{\xi}\imath_{\bar{\xi}}H
$$
where $\xi=\xi^mK_m\in TG$ and $\eta=\eta_m\ell^m\in T^*G$ and
$$
[K_m,K_n]=f_{mn}{}^pK_p \qquad  \d\ell^m+\frac{1}{2}f_{np}{}^m\ell^n\wedge\ell^p=0
$$
The $H$-flux is given by
$$
H=\frac{1}{6}H_{mnp}\ell^m\wedge\ell^n\wedge\ell^p
$$
where $H_{mnp}$ is a constant The requirement that $\d H=0$ leads to and algebraic Bianchi identity $H_{[mn|t}f_{|pq]}{}^t=0$ which determines
what $H$-flux is permissible on the group manifold\footnote{For semi-simple groups $G$ there is an invertible Cartan-Killing metric
$\eta_{mn}=\frac{1}{2}f_{mp}{}^qf_{nq}{}^p$ which may be used to raise and lower the Lie algebra indices of $G$. An example of a permissible flux
is given by the standard WZNW models on $G$, for which $H_{mnp}=\eta_{mq}f_{np}{}^q$).}. Putting these tangent and cotangent elements into the
$H$-twisted Courant bracket gives
\begin{eqnarray}
\left[Y,\bar{Y}\right]_H&=&\xi^m\bar{\xi}^n\left(\left[K_m,K_n\right]+\imath_m\imath_nH\right)+\xi^m\bar{\eta}_n\left({\cal
L}_m\ell^n-\frac{1}{2}\d(\imath_m\ell^n)\right)-\eta_m\bar{\xi}^n\left({\cal
L}_n\ell^m+\frac{1}{2}\d(\imath_n\ell^m)\right)\nonumber\\
&=&\xi^m\bar{\xi}^n\left(f_{mn}{}^pK_p+H_{mnp}\ell^p\right)-\xi^m\bar{\eta}_nf_{mp}{}^n\ell^p+\eta_m\bar{\xi}^nf_{np}{}^m\ell^p\nonumber
\end{eqnarray}
Comparing coefficients of $Y$ and $\bar{Y}$ gives the algebra
$$
\left[K_m,K_n\right]_H=f_{mn}{}^pK_p+H_{mnp}\ell^p\qquad \left[\ell^m,K_n\right]_H=f_{np}{}^m\ell^p\qquad \left[\ell^m,\ell^n\right]_H=0
$$
How does this relate to the twisted Drinfel'd double $\mathfrak{h}_H$? A left-Invariant representation of $\mathfrak{h}_H$ is
$$
\left[Z_m,Z_n\right]=f_{mn}{}^pZ_p+H_{mnp}X^p\qquad \left[X^m,Z_n\right]=f_{np}{}^mX^p\qquad \left[X^m,X^n\right]=0
$$
These vectors $T_M$ of the doubled group formalism can be written as generalised vectors in $TG\oplus T^*G$ by imposing the constraint $({\cal
P}^M|{\cal P}^N)=L^{MN}$ \cite{Dall'Agata:2007sr}. The constraint imposes the identification of $\ell^m$ with $X^m$. Restricting $Z_m$ to the
coset $G=\cG/\widetilde{G}_L$ recovers $K_m$ (i.e. $Z_m|_{G}=K_m$), and the identification of $\mathfrak{h}_H$ with the twisted
Courant bracket on $TG\oplus T^*G$ is complete. For the case where $H_{mnp}=0$ the Drinfel'd double is isomorphic to the Courant bracket as
described above. The addition of $H$-flux `twists' the Drinfel'd double in such a way that the new Lie-algebra is isomorphic to the
$H$-twisted Courant bracket.

It is interesting to note that a similar result, a truncation of the theory to a $\tilde{x}_i$ independent sector being associated to the Courant bracket, has been obtained from a doubled field theory \cite{HullZwiebach1,HullZwiebach2}.  More generally we might expect that the algebra (\ref{****}) has a natural description in terms of an $H$- and $R$-twisting of the bi-algebra (\ref{a}). It would be interesting to see how this general bracket relates to the C-bracket discovered in \cite{HullZwiebach1,HullZwiebach2}.

\section{World-sheet theories for doubled geometries}

In this section we shall consider two world-sheet theories describing the embedding of the world-sheet into the geometries discussed in the previous section. The doubled formalism introduced in \cite{Hull ``A geometry for non-geometric string backgrounds''} has proven to be an exceptionally useful tool in elucidating the structure of non-geometric backgrounds in
string theory. Despite this success many issues remain unclear and the scope of applicability of the formalism of \cite{Hull ``A geometry for
non-geometric string backgrounds''} is limited to those backgrounds which can be understood as torus fibrations and are necessarily locally
geometric. The standard world-sheet formulation of string theory is best adapted to studying globally geometric backgrounds. The formalism of
\cite{Hull ``A geometry for non-geometric string backgrounds''} extends this to include locally geometric backgrounds which are globally
non-geometric, such as T-folds. In \cite{Hull:2007jy,HullRRE} it was shown that a more powerful formalism can be constructed in which the doubled
background is locally a group manifold. A key point is that the doubled group will generally not be a product of tori or even a torus fibration
but will have a more general, non-abelian, structure. This doubled group perspective clarifies the nature of the geometry underlying T-duality
and overcomes many of the misleading features of the doubled torus construction. The Drinfel'd double is an example of a Lie-algebra which
generates such a doubled group.

\subsection{Hamiltonian construction of the world-sheet theory}

We consider here the sigma model, whose target space is locally the Poisson-Lie group, proposed by Klim\v{c}ik and Severa in
\cite{Klimcik:1995ux} and recast it in the language of \cite{HullRRE}. The global structure of the target space is taken to be that of a doubled
twisted torus. The re-expression of the Klim\v{c}ik-Severa model allows for three advancements: Firstly, contact can be made with older attempts
to construct string sigma-models with manifest T-duality symmetries \cite{Tseytlin:1990va}. Secondly, and most importantly, the proposed
Poisson-Lie duality is placed firmly within the context of the study of non-geometric backgrounds and thirdly, it leads to a generalization of
the considerations of both Hull \cite{Hull ``A geometry for non-geometric string backgrounds''} and Klim\v{c}ik and Severa \cite{Klimcik:1995dy}
to produce a world-sheet formulation of the target space results demonstrated in \cite{Hull:2007jy} analogous to that introduced in \cite{HullRRE}. Such sigma models have also been studied in this context in \cite{Dall'Agata:2008qz,Avramis:2009xi}.

The first task is to construct a manifestly T-duality invariant sigma-model for $M_d\times T^D$ where the explicit dependence
on the coordinates of $M_d$ will be neglected as it is the part of the sigma model which describes the embedding into $T^D$ which is of
prime interest; however, a full description of the theory requires also including the embedding into $M_d$.\footnote{In a T-duality context,
the coordinates $y^{\mu}$ on $M_d$ are often referred to as spectator coordinates as they do not correspond to directions that are
dualised along. The inclusion of the spectator coordinates in a doubled sigma model was given in \cite{RRE}.} The starting point is the action for the bosonic string
\begin{equation}\label{standard lagrangian}
S=\frac{1}{4\pi\alpha'}\oint_{\Sigma}\d^2\sigma\sqrt{h}h^{\alpha\beta}g_{ij}\partial_{\alpha}x^i\partial_{\beta}x^j+
\frac{1}{4\pi\alpha'}\oint_{\Sigma}\d^2\sigma\varepsilon^{\alpha\beta}B_{ij}\partial_{\alpha}x^i\partial_{\beta}x^j+ \frac{1}{4\pi}\oint_{\Sigma}\d^2\sigma\sqrt{h}\phi R(h)
\end{equation}
We shall choose to work in the gauge in which the world-sheet metric is $h=diag\{1,-1\}$ which sets the two-dimensional curvature to zero
$R(h)=0$ and $\varepsilon^{\tau\sigma}=-\varepsilon^{\sigma\tau}=1$. The dilaton term, which vanishes in this gauge, has no $\alpha'$ dependence
and does not contribute directly to the discussion of T-duality\footnote{That there is no incompatibility with the statements that the dilaton
transforms
 under T-duality \cite{Buscher ``A Symmetry of the String Background Field Equations'',Giveon:1993ai} and that T-duality
leaves the string coupling invariant was explained in
\cite{Alvarez:1996vt}.}. It is convenient to set $2\pi\alpha'=1$ and to give the world-sheet theory in terms of the Lagrangian
$$
\mathscr{L}=\frac{1}{2}g_{ij}\left(\partial_{\tau}x^i\partial_{\tau}x^j
-\partial_{\sigma}x^i\partial_{\sigma}x^j\right)+B_{ij}\partial_{\tau}x^i\partial_{\sigma}x^j
$$
The target space of interest is $T^D$ so $g_{ij}$ and $B_{ij}$ are taken to be independent of the embedding $x:\Sigma\rightarrow T^D$ but may
depend on the coordinates $y^{\mu}$ on $M_d$. In fact, throughout this paper, the lower case $g_{ij}$ and $B_{ij}$ will be used to denote a
background which is independent of all but the spectator coordinates. T-duality is a symplectomorphism of the phase space of the world-sheet
theory (a canonical transformation) and as such can be thought of as a basis transformation in the world-sheet phase space. The goal is to recast
the world-sheet Lagrangian such that the T-duality symmetry is manifest, or put another way, written in a form in which the symmetries of the
string phase space are more apparent. The canonical momentum $\mu_i$, conjugate to $x^i$, is
$$
\mu_i=\frac{\partial\mathscr{L}}{\partial(\partial_{\tau}x^i)}=g_{ij}\partial_{\tau}x^j+B_{ij}\partial_{\sigma}x^j
$$
The Hamiltonian density $\mathscr{H}=\mu_i\partial_{\tau}x^i-\mathscr{L}$ can be written in a manifestly T-duality invariant form
$$
\mathscr{H}=\frac{1}{2}\Psi^I{\cal M}_{IJ}\Psi^J
$$
where
\begin{equation}\label{M}
\Psi^I=\left(
        \begin{array}{c}
          \mu_i \\
          \partial_{\sigma}x^i \\
        \end{array}
      \right)
 \qquad  {\cal M}_{IJ}=\left(%
\begin{array}{cc}
  g_{ij}+B_{ik}g^{kl}B_{lj} & g^{jk}B_{ki} \\
  g^{ik}B_{kj} & g^{ij} \\
\end{array}%
\right)
\end{equation}
That the Hamiltonian density may be written in such a
duality invariant form should be of no surprise since T-duality, as remarked above, is a canonical transformation \cite{Sfetsos:1997pi,Lozano:1995jx,Lozano:1996sc}.
The Lagrangian may therefore be written as
$$
\mathscr{L}=\mu_i\partial_{\tau}x^i-\mathscr{H}
$$
In \cite{Tseytlin:1990va} it was proposed that the momentum be associated with a dual coordinate $\tilde{x}_i$ in the following way
$$
\mu_i=\partial_{\sigma}\tilde{x}_i
$$
For example, the $\mu_i\partial_{\tau}x^i$ term becomes $\partial_{\sigma}\tilde{x}_i\partial_{\tau}x^i$. The degrees of freedom are formally
doubled and doubled coordinates $\mathbb{X}^I=(x^i,\tilde{x}_i)$ are introduced. This lifts the canonical transformation to the status of a
geometric transformation on the doubled coordinates $\mathbb{X}^I$. The Lagrangian may then be written
\begin{equation}\label{tseytlin doubled lagrangian}
\mathscr{L}=\frac{1}{2}L_{IJ}\partial_{\tau}\mathbb{X}^I\partial_{\sigma}\mathbb{X}^J-\frac{1}{2}{\cal
M}_{IJ}\partial_{\sigma}\mathbb{X}^I\partial_{\sigma}\mathbb{X}^J
+\frac{1}{2}\Omega_{IJ}\partial_{\tau}\mathbb{X}^I\partial_{\sigma}\mathbb{X}^I
\end{equation}
where
$$
L_{IJ}=\left(%
\begin{array}{cc}
  0 & \delta^i{}_j \\
  \delta_i{}^j & 0 \\
\end{array}%
\right) \qquad  \Omega_{IJ}=\left(%
\begin{array}{cc}
  0 & \delta^i{}_j \\
  -\delta_i{}^j & 0 \\
\end{array}%
\right)\nonumber
$$
The first and last terms together give the $\mu_i\partial_{\sigma}X^i$ term. One can think of this as a doubled formalism in which world-sheet
Lorentz invariance is not manifest \cite{Tseytlin:1990va}. As a final comment, note that the momentum may be identified with the generator
$\mu_i\sim\partial_i$ and so that the isomorphism between $(T\oplus T^*)(T^D)$ and $T(T^{2D})$ that leads to the identification $d{\ti
x}_i=\partial_i$ coming from $(d\mathbb{X}^I|d\mathbb{X}^J)=L^{IJ}$ is quite natural.

\subsubsection{Sigma model on the doubled group}

In \cite{Klimcik:1995dy,Klimcik:1995ux} a sigma model describing the embedding of the worldsheet $\Sigma$ into a doubled group $\cG$ was
proposed (again, the explicit dependence on the spectator fields $y^{\mu}$ are suppressed, although the metric $\mathcal{M}$ and product
structure ${\cal R}$ may depend on $y^{\mu}$)
\begin{eqnarray}\label{Klimcik action}
S&=&\frac{1}{2}\oint_{\Sigma}\d^2\sigma\langle
h^{-1}\partial_{\sigma}h|h^{-1}\partial_{\tau}h\rangle-\frac{1}{2}\oint_{\Sigma}\d^2\sigma\langle h^{-1}\partial_{\sigma}h|{\cal
R}|h^{-1}\partial_{\sigma}h\rangle\nonumber\\
 &&+\frac{1}{12}\int_{V}\d^3\sigma'\varepsilon^{\alpha'\beta'\gamma'}\langle h^{-1}\partial_{\alpha'}h|[h^{-1}\partial_{\beta'}h,h^{-1}\partial_{\gamma'}h]\rangle
\end{eqnarray}
where $h\in\cG$ and $\alpha'$, $\beta'$ and $\gamma'$ label coordinates $\sigma'$ on $V$ where $\partial V=\Sigma$. Using the notation
$h^{-1}\partial_{\alpha}h={\cal P}_{\alpha}{}^MT_M$, introduced in (\ref{doubled forms}) where now ${\cal P}$ refers to the pull
back of the left-invariant forms on $\cG$ to the world-sheet $\Sigma$, this sigma model may be written as
\begin{eqnarray}
S&=&\frac{1}{2}\oint_{\Sigma}d^2\sigma\langle T_M|T_N\rangle{\cal P}_{\sigma}{}^M{\cal
P}_{\tau}{}^N-\frac{1}{2}\oint_{\Sigma}d^2\sigma\langle T_M|{\cal R}|T_N\rangle{\cal P}_{\sigma}{}^M{\cal
P}_{\sigma}{}^N\nonumber\\
&&+\frac{1}{12}\int_{V}\langle T_M|[T_N,T_P]\rangle{\cal P}^M\wedge {\cal P}^N\wedge {\cal P}^P\nonumber
\end{eqnarray}
Furthermore, it is shown in the Appendix that
$$
{\cal M}_{MN}=\langle T_M|{\cal R}|T_N\rangle   \qquad  [T_M,T_N]=t_{MN}{}^PT_P
$$
The doubled action may then be written as\footnote{One should really write the Wess-Zumino term as $\frac{1}{12}\int_Vt_{MNP}\Phi^M\wedge
\Phi^N\wedge\Phi^P$ where $\Phi^M$ depends on the coordinates $(\sigma'_1,\sigma'_2,\sigma'_3)$ on $V$ such that
$\Phi^M(\sigma'_1,\sigma'_2,\sigma'_3)|_{\Sigma}={\cal P}^M(\tau,\sigma)$. We shall refer to both the pull-back of the one-forms in $T^*\cX$ to
$\Sigma$ and $V$ both as ${\cal P}$. No confusion should result from this abuse of notation as any quantity defined on $V$ will always appear
under the integral $\int_V$. The classical physics only depends on the fields ${\cal P}^M$ defined as pull-backs to $\Sigma$.}
\begin{equation}\label{doubled hamiltonian action}
S=\frac{1}{2}\oint_{\Sigma}\d^2\sigma L_{MN}{\cal P}_{\sigma}{}^M{\cal P}_{\tau}{}^N-\frac{1}{2}\oint_{\Sigma}\d^2\sigma{\cal M}_{MN}{\cal
P}_{\sigma}{}^M{\cal P}_{\sigma}{}^N +\frac{1}{12}\int_{V}t_{MNP}{\cal P}^M\wedge {\cal P}^N\wedge {\cal P}^P
\end{equation}
This action is the non-abelian generalisation of (\ref{tseytlin doubled lagrangian}). The benefit of this formulation of the theory over the standard formulation (\ref{standard lagrangian}) is two-fold. Firstly the action of
$O(D,D)$ and hence the Poisson-Lie map is manifest and linear. Secondly, the action allows more general backgrounds to be understood much more
simply than the starting point (\ref{standard lagrangian}). In particular this sigma model can
describe embeddings of the world-sheet into backgrounds with $H$-flux and target spaces which are not even locally geometric. A clear drawback of this
construction is that Lorentz invariance on the world-sheet is not manifest. Doubled sigma models with manifest world-sheet Lorenz invariance were introduced in \cite{HullRRE} (see also \cite{RRE}) and will be discussed in the present context towards the end of this section.

\noindent\textbf{Recovering the conventional description}

The sigma model (\ref{doubled hamiltonian action}) on $\cX$ has a manifest rigid left-acting symmetry $\cG_L$. A sigma model on the coset $\cX/\widetilde{G}_L$ is
constructed by gauging the left-acting sub-group $\widetilde{G}_L$ in the doubled sigma-model above\footnote{The generators of the left-action
$\widetilde{T}_M$ are right-invariant and generally will not be globally defined on $\cX=\cG/\G_{\cG}$. In general, the
generator $(\widetilde{X}^m)_a$ of $\widetilde{G}_L$ in a local coordinate patch $\mathscr{U}_a$ will be related to $(\widetilde{X}^m)_b$ in a
patch $\mathscr{U}_b$ by $(\widetilde{X}^m)_a=(\gamma^m{}_n)_{ab}(\widetilde{X}^n)_b$ where $\gamma_{ab}\in\G_{\cG}$ defined on the
overlap $\mathscr{U}_a\cap \mathscr{U}_b$. This issue will not be considered further here but is discussed at length in \cite{HullRRE}.}. The
obstructions to gauging sigma models with Wess-Zumino terms are discussed in \cite{Hull and Spence} and the various criteria a sub-group must
satisfy in order for the gauged theory to also encode two-dimensional physics were clarified. It can be shown that these criteria reduce, in
this case, to the statement that the gauge group must be a null sub-group with respect to $L_{MN}$, in other words that it must be maximally
isotropic \cite{HullRRE}. The maximally isotropic sub-group we choose to
gauge is $\widetilde{G}_L$. This is done by introducing $\tilde{\mathfrak{g}}$-valued world-sheet one-forms $C=C_{\sigma}\d\sigma$ where
$C_{\sigma}=C_{\sigma m}\widetilde{T}^m$ which transform under the left action $\widetilde{G}_L$ as
$$
\delta C_{\sigma m}=\partial_{\sigma}\varepsilon_m+f_m{}^{np}\varepsilon_pC_{\sigma n}
$$
where $\varepsilon$ is a gauge parameter which depends on the world-sheet coordinate $\sigma$. It will be seen that the sigma model is manifestly
invariant under $\tau$-dependent gauge transformations acting from the left so there is no need to introduce one-forms
$C_{\tau}\d\tau$ explicitly. It is useful to define
$$
{\cal C}_{\sigma}=h^{-1}C_{\sigma}h
$$
and the minimal coupling terms $\widehat{{\cal P}}_{\sigma}=h^{-1}(\partial_{\sigma}+C_{\sigma})h$ as
$$
\widehat{{\cal P}}_{\sigma}{}^M={\cal P}_{\sigma}{}^M+{\cal C}_{\sigma}{}^M
$$
The gauging proceeds by minimally coupling of the first two terms in (\ref{doubled hamiltonian action}) by replacing ${\cal P}_{\sigma}{}^M$
with $\widehat{{\cal P}}_{\sigma}{}^M$. An important result of \cite{Hull and Spence} is that the gauged Wess-Zumino term may be written as
$$
\frac{1}{12}\int_{V}t_{MNP}\widehat{{\cal P}}^M\wedge \widehat{{\cal P}}^N\wedge \widehat{{\cal P}}^P = \frac{1}{2}\oint_{\Sigma}L_{MN}{\cal
P}^M\wedge {\cal C}^N +\frac{1}{12}\int_{V}t_{MNP}{\cal P}^M\wedge {\cal P}^N\wedge {\cal P}^P
$$
The minimal coupling of the term $\frac{1}{2}L_{MN}{\cal P}_{\sigma}{}^M{\cal P}_{\tau}{}^N\rightarrow\frac{1}{2}L_{MN}\widehat{{\cal
P}}_{\sigma}{}^M{\cal P}_{\tau}{}^N$ introduces another factor of $\frac{1}{2}\oint_{\Sigma}L_{MN}{\cal P}_{\tau}{}^M{\cal C}_{\sigma}{}^N$.
Finally, the $\widetilde{G}_L$-gauged sigma model on $\cG$, which is equivalent to a sigma model on $\cG/\widetilde{G}_L$, may be
written as
\begin{eqnarray}\label{gauged}
S&=&\frac{1}{2}\oint_{\Sigma}\d^2\sigma L_{MN}{\cal P}_{\sigma}{}^M{\cal P}_{\tau}{}^N-\frac{1}{2}\oint_{\Sigma}\d^2\sigma{\cal
M}_{MN}\widehat{{\cal P}}_{\sigma}{}^M\widehat{{\cal P}}_{\sigma}{}^N +\oint_{\Sigma}\d^2\sigma L_{MN}{\cal P}_{\tau}{}^M{\cal
C}_{\sigma}{}^N\nonumber\\
 &&+\frac{1}{12}\int_{V}t_{MNP}{\cal P}^M\wedge {\cal P}^N\wedge {\cal P}^P
\end{eqnarray}
This can also be written as
$$
S=\frac{1}{2}\oint_{\Sigma}\d^2\sigma\langle {\cal P}_{\sigma}|{\cal P}_{\tau}\rangle -\frac{1}{2}\oint_{\Sigma}\d^2\sigma\langle \widehat{{\cal
P}}_{\sigma}|{\cal R}|\widehat{{\cal P}}_{\sigma}\rangle +\oint_{\Sigma}\d^2\sigma\langle{\cal P}_{\tau}|{\cal C}_{\sigma}\rangle
+\frac{1}{12}\int_{V}\d^3\sigma'\varepsilon^{\alpha'\beta'\gamma'}\langle {\cal P}_{\alpha'}|[{\cal P}_{\beta'},{\cal P}_{\gamma'}]\rangle
$$
Although this last form of the gauged action may appear to be independent of the choice of Lie algebra basis, it is only really independent of
the of basis changes which preserve the choice of Manin triple. The one-form $C_{\sigma}$ (and $\mathcal{C}_{\sigma}$), is only
specified once a Manin triple has been chosen and a subgroup is chosen to be gauged.

\noindent\textbf{Sigma models on $\cG/\widetilde{G}_L$ and $\cX/\widetilde{G}_L$}

Let us now see how the gauged sigma model above reduces to the standrd one on $G$. Consider the Manin triple $(\mathfrak{h},\widetilde{\mathfrak{g}},\mathfrak{g})$ and the corresponding decomposition of the elements of
$\cG$ as $h=\tilde{g}g$ where $\tilde{g}\in \widetilde{G}$ and $g\in G$. The sigma model of interest is the gauging of the sigma model
on the doubled group $\cG$, or more generally the twisted torus $\cX$, by the introduction of the one-form
$C_{\sigma}=C_{\sigma m}\widetilde{T}^m$.

In terms of the decomposition $h=\tilde{g}g$, where the left action is denoted by $\widetilde{G}_L:h\rightarrow \xi h$ where $\xi\in
\widetilde{G}$. $\widetilde{G}_L$ acts as
$$
\widetilde{G}_L:\tilde{g}\rightarrow \xi\tilde{g}   \qquad  \widetilde{G}_L:g\rightarrow g
$$
The minimally coupled left-invariant $\widehat{\mathcal{P}}_{\sigma}=h^{-1}(\partial_{\sigma}+C_{\sigma})h$ may be written in terms of the maximally isotropic generators $(T_m,\widetilde{T}_m)$ as
 $\widehat{\mathcal{P}}^M=\widehat{P}^mT_m+\widehat{Q}_m\widetilde{T}^m$, where
$$
\widehat{P}^m=(g^{-1}\d g)^m-\beta^{mn}(\tilde{g}^{-1}D\tilde{g})_n \qquad \widehat{Q}_m=A_m{}^n(\tilde{g}^{-1}D\tilde{g})_n
$$
and $D=\d+C_{\sigma}\d\sigma$. It is useful to write $\widehat{{\cal P}}^M$ as $\widehat{{\cal P}}^M=\widehat{\Phi}^N{\cal
V}_N{}^M $ where
$$
\widehat{\Phi}^M=\left(%
\begin{array}{cc}
  \ell^m & \widehat{Q}_m \\
\end{array}%
\right) \qquad {\cal
V}=\left(%
\begin{array}{cc}
  \delta_n{}^m & 0 \\
  -\pi^{mn} & \delta^n{}_m \\
\end{array}%
\right)
$$
and $\ell^m=(g^{-1}\d g)^m$ is the left-invariant one-form on $T^*G$. The matrix $\mathcal{V}$ acts as a $\beta$-shift so that ${\cal V}{\cal
M}(E){\cal V}^T={\cal M}(\cal E)$ where
\begin{equation}\label{E}
{\cal E}=(E^{-1}+\pi)^{-1}
\end{equation}
The second term in the action then becomes
$$
-\frac{1}{2}\oint_{\Sigma}\d^2\sigma{\cal M}_{MN}(E)\widehat{{\cal P}}_{\sigma}{}^M\widehat{{\cal
P}}_{\sigma}{}^N=-\frac{1}{2}\oint_{\Sigma}\d^2\sigma{\cal M}_{MN}({\cal E})\Phi_{\sigma}{}^M\Phi_{\sigma}{}^N
$$
This term may then be expanded out as
$$
{\cal M}_{MN}(E)\widehat{{\cal P}}_{\sigma}{}^M\widehat{{\cal P}}_{\sigma}{}^N=G^{mn}\widehat{Q}_{\sigma m}\widehat{Q}_{\sigma
n}-2G^{mp}B_{pn}\widehat{Q}_{\sigma m}\ell^n{}_{\sigma}+(G_{mn}+B_{mp}G^{pq}B_{nq})\ell^m{}_{\sigma}\ell^n{}_{\sigma}
$$
where\footnote{This is in contrast to $g_{mn}=E_{(mn)}$ and $b_{mn}=E_{[mn]}$. Note that $G_{mn}(e)=g_{mn}$ and $B_{mn}(e)=b_{mn}$ since $\pi^{mn}(e)=0$.} $G_{mn}(g)={\cal E}_{(mn)}$ and $B_{mn}(g)={\cal E}_{[mn]}$.

Consider now the Wess-Zumino term of the sigma model for this choice of Manin triple. It is convenient to write the left-invariant one forms as
\begin{equation}
{\cal P}=g^{-1}(r+\tilde{\ell})g=g^{-1}\mathscr{P} g
\end{equation}
where $r=\d gg^{-1}$ and $\tilde{\ell}=\tilde{h}^{-1}\d\tilde{h}$. Using the fact that the inner product $\langle\,|\,\rangle$ is adjoint-invariant
so that $\langle{\cal P}_{\alpha'}| [{\cal P}_{\beta'}, {\cal P}_{\gamma'}]\rangle =\langle\mathscr{P}_{\alpha'}|[\mathscr{P}_{\beta'},
\mathscr{P}_{\gamma'}]\rangle$ and the Wess-Zumino term may be written as
$$
S_{wz}=-\frac{1}{4}\int_V\left(f_{mn}{}^pr^m\wedge r^n\wedge \tilde{\ell}_p+c_m{}^{np}r^m\wedge \tilde{\ell}_n\wedge \tilde{\ell}_p\right)
=\frac{1}{2}\oint_{\Sigma}r^m\wedge \tilde{\ell}_m
$$
which is the generalisation of the form $\Omega_{IJ}\d\mathbb{X}^I\wedge \d\mathbb{X}^J$ seen in the abelian case (\ref{tseytlin doubled
lagrangian}). Finally, the Wess-Zumino term can be written as
$$
\frac{1}{2}\oint_{\Sigma}\d^2\sigma\langle \d gg^{-1}|{\ti g}^{-1}\d{\ti g}\rangle =\frac{1}{2}\oint_{\Sigma}\d^2\sigma\langle \ell|Q\rangle
=\frac{1}{2}\oint_{\Sigma}\d^2\sigma\left(\ell^m{}_{\tau}Q_{\sigma m}-\ell^m{}_{\sigma}Q_{\tau m}\right)
$$
The first term in the action may also be simplified
$$
\frac{1}{2}\oint_{\Sigma}\d^2\sigma L_{MN}{\cal P}_{\sigma}{}^M{\cal P}_{\tau}{}^N
=\frac{1}{2}\oint_{\Sigma}\d^2\sigma\left(\ell^m{}_{\tau}Q_{\sigma m} +\ell^m{}_{\sigma}Q_{\tau m}\right)
$$
where the $O(D,D)$ invariance of $L_{MN}$ has been used, $L_{PQ}\mathcal{O}^P{}_M\mathcal{O}^Q{}_N=L_{MN}$. The extra term $\langle{\cal
P}_{\tau}|{\cal C}_{\sigma}\rangle$ introduced in the gauged theory can be written, using the adjoint invariance of the inner product, as
$$
\langle{\cal P}_{\tau}|{\cal C}_{\sigma}\rangle =\langle\partial_{\tau}hh^{-1}|\mathcal{C}_{\sigma}\rangle =\langle{\ti
g}^{-1}\partial_{\tau}{\ti g}+\partial_{\tau}gg^{-1}|{\ti g}^{-1}\mathcal{C}_{\sigma}{\ti g}\rangle= \langle\ell_{\tau}|{\ti
g}^{-1}C_{\sigma}{\ti g}\rangle
$$
so that
$$
\frac{1}{2}\oint_{\Sigma}d^2\sigma L_{MN}{\cal P}_{\sigma}{}^M{\cal P}_{\tau}{}^N +\oint_{\Sigma}d^2\sigma L_{MN}{\cal P}_{\tau}{}^M{\cal
C}_{\sigma}{}^N +\frac{1}{12}\int_{V}t_{MNP}{\cal P}^M\wedge {\cal P}^N\wedge {\cal P}^P=
\oint_{\Sigma}d^2\sigma\ell_{\tau}{}^m\widehat{Q}_{\sigma m}
$$
Finally, the action (\ref{gauged}) can be written in terms of Lagrangian
$$
\mathscr{L}=-\frac{1}{2}G^{mn}\widehat{Q}_{\sigma m}\widehat{Q}_{\sigma n}+G^{mp}B_{pn}\widehat{Q}_{\sigma
m}\ell_{\sigma}{}^n-\frac{1}{2}(G_{mn}+B_{mp}G^{pq}B_{nq})\ell^m{}_{\sigma}\ell_{\sigma}{}^n+\ell_{\tau}{}^m\widehat{Q}_{\sigma m}
$$
Note that only $Q_{\sigma}=g^{-1}({\ti g}^{-1}\partial_{\sigma}{\ti g})g$ appears in the Lagrangian, $Q_{\tau}$ does not, so that the
theory is invariant under arbitrary left-acting $\tau$-dependent gauge transformations ${\ti g}\rightarrow \xi(\tau){\ti g}$ as claimed above.
Completing the square in $\widehat{Q}_{\sigma m}$ gives
\begin{equation}\label{lagrangian}
\mathscr{L}={\cal E}_{mn}(g^{-1}\partial_-g)^m(g^{-1}\partial_+g)^n -\frac{1}{2}G^{mn}\lambda_m\lambda_n
\end{equation}
where $\partial_{\pm}=\partial_{\tau}\pm\partial_{\sigma}$ and
$$
\lambda_m=(A^{-1})_m{}^n\widetilde{A}_n{}^pC_{\sigma p}+Q_{\sigma m}-G_{mn}\ell_{\tau}{}^n-B_{mn}\ell_{\sigma}{}^n
$$
Integrating out the $C_{\sigma m}$ gives a contribution of
$\det[(A^{-1})\widetilde{A}G^{-1}\widetilde{A}^T(A^{-1})^T]=\det{(G^{-1})}$ to the path integral. This gives the
required correction to the dilaton and (\ref{lagrangian}) gives the expected Lagrangian on the Poisson-Lie group $G$ given in
\cite{Klimcik:1995dy}.

\noindent\textbf{Sigma models on $\cX/G_L$ and $\cG/G_L$}

Now consider the complementary Manin triple $(\mathfrak{h},\mathfrak{g},\widetilde{\mathfrak{g}})$ where the roles of $\mathfrak{g}$ and
$\widetilde{\mathfrak{g}}$ have been exchanged. An element of the doubled group may be written as $ h=g{\ti g}$ where $g\in G$ and
$\tilde{g}\in\widetilde{G}$. The left action $G_L$ acts as
$$
G_L:g\rightarrow \xi g  \qquad  G_L:\tilde{g}\rightarrow \tilde{g}
$$
The corresponding exchange of $T_m$ with $\widetilde{T}^m$ - an action of $L_{MN}$ on the space of generators $T_M\rightarrow
T^M=L^{MN}T_N$ - relates this decomposition to the decomposition $h={\ti g}g$ used above. One may think of this as
exchanging the the Manin triple $(\mathfrak{h},\mathfrak{g},\widetilde{\mathfrak{g}})$ with
$(\mathfrak{h},\widetilde{\mathfrak{g}},\mathfrak{g})$, or in the language of \cite{Hull ``A geometry for non-geometric string backgrounds''} as
choosing a different polarization of the Lie algebra related to the first by an action of the element $L_{MN}\in O(D,D;\Z)$. In order to gauge the subgroup $G_L$ the one-forms
$C_{\sigma}=C_{\sigma}{}^mT_m$ and ${\cal C}_{\sigma}=h^{-1}C_{\sigma}h$
must be introduced, which  transform under $G_L$ as
$$
\delta C_{\sigma}{}^m=\partial_{\sigma}\varepsilon^m+f_{np}{}^m\varepsilon^pC_{\sigma}{}^n
$$
It must be stressed that this is a-priori a different gauging than that considered above and so the gauged sigma models represent a-priori
different theories. One cannot, at this stage, infer that the sigma models on $\cX/\widetilde{G}_L$ and $\cX/G_L$ describe the
same physics.

The components of the minimally coupled left-invariant one-forms $\widehat{{\cal P}}^M$ are
$$
\widehat{P}^m=\widetilde{A}^m{}_n(g^{-1}Dg)^n \qquad  \widehat{Q}_m=({\ti g}^{-1}\d{\ti g})_m-\tilde{\beta}_{mn}(g^{-1}Dg)^n
$$
It is useful to write $\widehat{{\cal P}}^M=\widetilde{\Phi}^N{\cal W}_N{}^M $ where
$$
\widetilde{\Phi}^M=\left(%
\begin{array}{cc}
  \widehat{P}^m & {\ti \ell}_m \\
\end{array}%
\right) \qquad {\cal W}=\left(%
\begin{array}{cc}
  \delta_n{}^m & -\tilde{\pi}_{mn} \\
  0 & \delta^n{}_m \\
\end{array}%
\right)
$$
where $\tilde{\ell}_m=(\tilde{g}^{-1}\d\tilde{g})_m$ does not depend on the one-forms $C_{\sigma}$ and $\tilde{\pi}$ is the
Poisson structure $\tilde{\pi}=\widetilde{\beta}\widetilde{A}^{-1}$. In this case the matrix ${\cal W}$ acts to produce a
$b$-shift on the background tensor ${\cal W}{\cal M}(E){\cal W}^T={\cal M}(\widetilde{{\cal E}}^{-1})$ where
\begin{equation}\label{E dual}
\widetilde{{\cal E}}=(E+{\ti \pi})^{-1}
\end{equation}
It is convenient to write the action in terms of the symmetric and antisymmetric components of $\widetilde{\cal E}$ using the fact that ${\cal
M}(\widetilde{{\cal E}}^{-1})={\cal M}^{-1}(\widetilde{{\cal E}})$. The second term may then be written out as
$$
-\frac{1}{2}\oint_{\Sigma}\d^2\sigma{\cal M}_{MN}(E)\widehat{{\cal P}}_{\sigma}{}^M\widehat{{\cal
P}}_{\sigma}{}^N=-\frac{1}{2}\oint_{\Sigma}\d^2\sigma{\cal M}^{MN}(\widetilde{\cal E})\widetilde{\Phi}_{\sigma
M}\widetilde{\Phi}_{\sigma N}
$$
This term may then be expanded out as
$$
{\cal M}^{MN}(\widetilde{{\cal
E}})\widetilde{\Phi}_{\sigma}{}_M\widetilde{\Phi}_{\sigma}{}_N=\tilde{G}_{mn}\widehat{P}^m{}_{\sigma}\widehat{P}^n{}_{\sigma}
-2\tilde{G}_{mp}\tilde{B}^{pn}\widehat{P}^m{}_{\sigma}{\ti \ell}_{\sigma n}+(\tilde{G}^{mn}+\tilde{B}^{mp}\tilde{G}_{pq}\tilde{B}^{nq}){\ti
\ell}_{\sigma m}{\ti \ell}_{\sigma n}
$$
where
\begin{eqnarray}
\mathcal{M}^{MN}(\widetilde{\mathcal{E}})=\left(%
\begin{array}{cc}
  \tilde{G}^{mn}+\tilde{B}^{mp}\tilde{G}_{pq}\tilde{B}^{qn} & \tilde{G}_{np}\tilde{B}_{pm} \\
  \tilde{G}_{mp}\tilde{B}^{pn} & \tilde{G}_{mn} \\
\end{array}%
\right) \qquad \tilde{G}^{mn}=\widetilde{{\cal E}}^{(mn)}  \qquad  \tilde{B}^{mn}=\widetilde{{\cal E}}^{[mn]}
\end{eqnarray}
The Wess-Zumino term may be written as
$$
\oint_{\Sigma}\frac{1}{2}\langle \d gg^{-1}|{\ti g}^{-1}d{\ti g}\rangle =\oint_{\Sigma}\frac{1}{2}\langle P|{\ti \ell}\rangle
=\frac{1}{2}\oint_{\Sigma}\d^2\sigma \left(P^m{}_{\tau}{\ti \ell}_{\sigma m}-P^m{}_{\sigma}{\ti \ell}_{\tau m}\right)
$$
so that
$$
\oint_{\Sigma}\d^2\sigma L_{MN}{\cal P}_{\tau}{}^M{\cal C}_{\sigma}{}^N+\frac{1}{2}\oint_{\Sigma}\d^2\sigma L_{MN}{\cal
P}_{\sigma}{}^M{\cal P}_{\tau}{}^N +\frac{1}{12}\int_{V}t_{MNP}{\cal P}^M\wedge {\cal P}^N\wedge {\cal P}^P=
\oint_{\Sigma}\d^2\sigma\widehat{P}^m{}_{\tau}{\ti \ell}_{\sigma m}
$$
Completing the square in $\widehat{P}^m{}_{\sigma}$ gives
\begin{equation}\label{Lagrangian dual}
\mathscr{L}=\widetilde{{\cal E}}^{mn}({\ti g}^{-1}\partial_-{\ti g})_m({\ti g}^{-1}\partial_+{\ti g})_n
-\frac{1}{2}\tilde{g}_{mn}\lambda_m\lambda_n
\end{equation}
where $\partial_{\pm}=\partial_{\tau}\pm\partial_{\sigma}$ and
$$
\lambda^m=\widehat{P}_{\sigma}{}^m-\tilde{G}^{mn}{\ti \ell}_{\tau n}-\tilde{B}^{mn}{\ti \ell}_{\sigma n}
$$
Integrating out $C_{\sigma}$ gives a $\det(\tilde{G}^{-1})$ contribution to the dilaton and the `dual' theory found in
\cite{Klimcik:1994gi} is recovered.

\subsection{Twisted Poisson-Lie Structures}

The considerations above address models in which the target space is a twisted torus $\cX=\cG/\G_{\cG}$ associated with
the Drinfel'd double
\begin{equation}
[T_m,T_n]=f_{mn}{}^pT_p  \qquad [\widetilde{T}^m,T_n]=f_{np}{}^m\widetilde{T}^p-c_n{}^{mp}T_p \qquad
[\widetilde{T}^m,\widetilde{T}^n]=c^{mn}{}_p\widetilde{T}^p
\end{equation}
where the two sub-algebras $G$ and $\widetilde{G}$, generated by $T_m$ and $\widetilde{T}^m$, are maximally isotropic with respect to the inner
product $\langle\,|\,\rangle$ which defines the $O(D,D)$ structure. If one drops the requirement that $T_m$ and $\widetilde{T}^m$ must generate
subalgebras and simply require that the sets of generators be null with respect to the inner product, and therefore be compatible with the
$O(D,D)$ structure, then more general doubled twisted tori, based on an algebras which are not Drinfel'd doubles, are allowed. In particular we
may allow target spaces $\cX=\cG/\G_{\cG}$ where the algebra generating $\cG$ is
\begin{equation}\label{twisted algebra}
[T_m,T_n]=f_{mn}{}^pT_p+H_{mnp}\widetilde{T}^p  \qquad [\widetilde{T}^m,T_n]=\gamma_{np}{}^m\widetilde{T}^p+h_n{}^{mp}T_p
\end{equation}
\begin{equation}
[\widetilde{T}^m,\widetilde{T}^n]=c^{mn}{}_p\widetilde{T}^p+R^{mnp}T_p
\end{equation}
If $R^{mnp}\neq 0$ ($H_{mnp}\neq 0$) then the set of generators $\widetilde{T}^m$ ($T_m$) do not close to form a sub-algebra. Such algebras may
be thought of as \emph{twisted Drinfel'd doubles}. The rationale behind this terminology that is that the extra structure constant $H_{mnp}$
plays the role of a flux in the background \cite{Kaloper ``The O(dd) story of massive supergravity'',Hull ``Flux compactifications of string
theory on twisted tori''} and in particular is responsible for $H$-twisting the Courant bracket in geometric $TG\oplus T^*G$ backgrounds
\cite{Gualtieri:2003dx,Gualtieri:2007ng}, where $G$ is locally a group manifold. Let us consider such backgrounds from the world-sheet perspective.

The doubled formalism (\ref{doubled hamiltonian action}) is written in terms of the pull-back of the left-invariant one-forms on $\mathfrak{h}$
$$
{\cal P}^M={\cal P}_{\tau}{}^M\d\tau+{\cal P}_{\sigma}{}^M\d\sigma
$$
to the world-sheet $\Sigma$. These forms obey the Bianchi identity
\begin{equation}\label{bianchi 2}
\partial_{\tau}{\cal P}_{\sigma}{}^M-\partial_{\sigma}{\cal P}_{\tau}{}^M=t_{NP}{}^M{\cal P}_{\tau}{}^N{\cal P}_{\sigma}{}^P
\end{equation}
where $t_{MN}{}^P$ are structure constants for $\mathfrak{h}$. This formalism is not restricted to Drinfel'd doubles, it could equally well be
employed to describe embeddings of the world-sheet into any group $\cG$ which is compatible with the $O(D,D)$ structure. In particular,
we are interested in generalizing the the application of this doubled sigma model to groups with algebras of the more general form (\ref{twisted
algebra}). As an example, consider the addition of $H$-flux to the Drinfel'd double corresponding to the group $\cG=G\ltimes \R^D$ (a group manifold with vanishing Poisson structure). The
Lie algebra of this $H$-twisted Drinfel'd double, which shall be denoted by $\mathfrak{h}_H$, is
$$
[T_m,T_n]=f_{mn}{}^pT_p+H_{mnp}\widetilde{T}^p\qquad [\widetilde{T}^m,T_n]=f_{np}{}^m\widetilde{T}^p\qquad [\widetilde{T}^m,\widetilde{T}^n]=0
$$
Now that the $T_m$ do not close to form a sub-algebra, the adjoint action $g^{-1}T_mg$ will in general include a contribution from the
$\widetilde{T}^m$ generators, so that $g^{-1}T_mg=A_m{}^nT_n+b_{mn}\widetilde{T}^n$, where $b_{mn}=\langle
g^{-1}T_mg|T_n\rangle$, so that the matrix (\ref{adjoint action}) takes the more general form
$$
g^{-1}\left(%
\begin{array}{c}
  T^m \\
  \widetilde{T}_m \\
\end{array}%
\right)g=\left(%
\begin{array}{cc}
  A^m{}_n & b_{mn} \\
  \beta^{mn} & (A^{-1})_m{}^n \\
\end{array}%
\right)\left(%
\begin{array}{c}
  T^n \\
  \widetilde{T}_n \\
\end{array}%
\right)
$$
The requirement $b_{mn}=-b_{nm}$ ensures that this matrix is an element of $O(D,D)$. For example, a torus $T^d$ with
constant flux
$$
H=\frac{1}{6}H_{mnp}dx^m\wedge dx^n\wedge dx^p
$$
where $H_{mnp}$ is a constant, gives $b_{mn}=\frac{1}{2}H_{mnp}X^p$. In order to recover the standard description of this background
the left-acting abelian symmetry generated by the right-invariant vectors $\widetilde{X}^m$ must be gauged. As discussed in the previous
section, the gauging of this left-action corresponds to minimal coupling of the $Q_{\sigma m}\rightarrow \widehat{Q}_{\sigma m}$. The
left-invariant one-forms on $\cX$ may be written as ${\cal P}^MT_M=P^mT_m+Q_m\widetilde{T}^m$ and, for the twisted double
$\mathfrak{h}_H$, these one-forms satisfy the Bianchi identities (\ref{bianchi 2}), which for the example $\mathfrak{h}_H$, considered here are
$$
\d P^m+\frac{1}{2}f_{np}{}^mP^n\wedge P^p=0   \qquad  \d Q_m+f_{mn}{}^pP^n\wedge Q_p+\frac{1}{2}H_{mnp}P^n\wedge P^p=0
$$
the left-invariant vector fields dual to these one-forms are a left-invariant generators for the above algebra. To recover the theory on
$\cX/\widetilde{G}_L$ one proceeds as before by gauging the rigid $G_L$ symmetry of the sigma model (\ref{doubled hamiltonian action}),
except now the $h$ take values in the group with Lie algebra $\mathfrak{h}_H$. The Wess-Zumino term for the action (\ref{doubled hamiltonian
action}) may be written as
$$
\frac{1}{12}\int_Vt_{MNP}{\cal P}^M\wedge{\cal P}^N\wedge{\cal P}^P=\frac{1}{2}\oint_{\Sigma}P^m\wedge Q_m+\frac{1}{6}\int_VH_{mnp}P^m\wedge
P^n\wedge P^p
$$
so that the gauged doubled action (\ref{gauged}) may then be written as
$$
S=-\frac{1}{2}\oint_{\Sigma}\d^2\sigma\,\,{\cal M}_{MN}(E)\widehat{{\cal P}}_{\sigma}{}^M\widehat{{\cal P}}_{\sigma}{}^N +\oint_{\Sigma}\d^2\sigma\,\,
P_{\tau}{}^m\widehat{Q}_{\sigma m} +\frac{1}{6}\int_VH_{mnp}P^m\wedge P^n\wedge P^p\nonumber
$$
Expanding out the first term
\begin{eqnarray}
S&=&\oint_{\Sigma}\d^2\sigma\,\,\left(\frac{1}{2}g^{mn}\widehat{Q}_{\sigma m}\widehat{Q}_{\sigma n}+g^{mp}B_{pn}\widehat{Q}_{\sigma m}P_{\sigma}{}^n
+\frac{1}{2}\left(g_{mn}+B_{mp}g^{pq}B_{qn}\right)P_{\sigma}{}^mP_{\sigma}{}^n +P_{\tau}{}^m\widehat{Q}_{\sigma
m}\right)\nonumber\\
&&+\frac{1}{6}\int_VH_{mnp}P^m\wedge P^n\wedge P^p\nonumber
\end{eqnarray}
and completing the square in $\widehat{Q}_{\sigma m}$ gives
$$
S=\frac{1}{2}\oint_{\Sigma}\d^2\sigma \,\, g^{mn}\lambda_m\lambda_n+\oint_{\Sigma}d^2\sigma\,\, E_{mn}P_{-}{}^mP_+{}^n +\frac{1}{6}\int_VH_{mnp}P^m\wedge P^n\wedge
P^p
$$
where $\lambda_m=\widehat{Q}_{\sigma m}-g_{mn}P_{\tau}{}^n-B_{mn}P_{\sigma}{}^n$. Integrating out the gauge fields gives a sigma model on
$G$ (or $G/\G_G$) with $G_L$ invariant $H$-flux as expected.

\subsection{Lagrangian construction of the world-sheet theory}

The doubled formalism (\ref{Klimcik action}) introduced in \cite{Klimcik:1995dy} has been shown to be applicable beyond its original remit and
can adequately describe string backgrounds which arise from general gaugings of $O(D,D)$ (as described in the introduction) not just those which
correspond to Poisson-Lie groups, but also those which include $H$ and $R$-flux. A regrettable feature of this formalism though is that manifest world-sheet Lorentz invariance is lost in the doubled action.
That such theories are implicitly Lorentz invariant, was investigated in \cite{Tseytlin:1990va} and can be seen directly in the manifest Lorentz
invariance of the gauged sigma models once the gauge fields have been integrated out. A doubled model in which Lorentz invariance is manifest at
all stages, would be more appealing and was introduced in \cite{HullRRE}.

The action describing the embedding of a closed string world-sheet $\Sigma$ into the target space $\cX$ is
\begin{eqnarray}
S&=&\frac{1}{4}\oint_{\Sigma}\d^2\sigma\,\,\sqrt{h}h^{\alpha\beta}{\cal H}_{IJ}\partial_{\alpha}\mathbb{X}^I\partial_{\beta}\mathbb{X}^J
+\frac{1}{12}\int_V\d^3\sigma'\,\,\varepsilon^{\alpha'\beta'\gamma'}{\cal K}_{IJK}\partial_{\alpha'}\mathbb{X}^I\partial_{\beta'}\mathbb{X}^J\partial_{\gamma'}\mathbb{X}^K
\nonumber\\
&&+\frac{1}{2\pi}\oint_{\Sigma}\d^2\sigma\,\,\sqrt{h}\phi R(h)\nonumber
\end{eqnarray}
where $V$ is an extension of the world-sheet such that $\partial V=\Sigma$ and
$$
{\cal H}_{IJ}(\mathbb{X},Y)={\cal M}_{MN}(Y)\mathcal{P}^M{}_I\mathcal{P}^N{}_J   \qquad
{\cal K}_{IJK}(\mathbb{X})=t_{MNP}\mathcal{P}^M{}_I\mathcal{P}^N{}_J\mathcal{P}^P{}_K
$$
In terms of the world-sheet Hodge star $*$ and wedge product $\wedge$, the action can be simply written as
\begin{eqnarray}\label{fibre action}
S=\frac{1}{4}\oint_{\Sigma}{\cal M}_{MN}{\cal P}^M\wedge *{\cal P}^N+\frac{1}{12}\int_Vt_{MNP}{\cal P}^M\wedge {\cal P}^N\wedge{\cal
P}^P+\frac{1}{2\pi}\oint_{\Sigma}\phi R(h)*1
\end{eqnarray}
${\cal P}=h^{-1}dh$ are the left-invariant one-forms, $\mathcal{M}$ is given by (\ref{M matrix}) and is taken to be independent of
$\mathbb{X}^I$ and $t_{MNP}=L_{MQ}t_{NP}{}^Q$. A gauge for the world-sheet metric such that $R(h)=0$ will be chosen. The left-invariant one-forms
satisfy the Bianchi identity
\begin{eqnarray}\label{bianchi}
\d{\cal P}^M+\frac{1}{2}t^M{}_{NP}{\cal P}^N\wedge{\cal P}^P&=&0
\end{eqnarray}
It is interesting to note that the action can be written in a way that is independent of a choice of Lie-algebra basis
$$
S=\frac{1}{4}\oint_{\Sigma}\d^2\sigma\sqrt{h}h^{\alpha\beta}\langle{\cal P_{\alpha}}|{\cal R}|{\cal P}_{\beta}\rangle
+\frac{1}{12}\int_V\d^3\sigma'\varepsilon^{\alpha'\beta'\gamma'}\langle{\cal P}_{\alpha'}|[{\cal P}_{\beta'},{\cal P}_{\gamma'}]\rangle
$$
The equations of motion are\footnote{A Lorentzian signature has been chosen for the world-sheet metric $h_{\alpha\beta}$ so that $*^2=+1$.
This is a convenience rather than a necessity and a signature such that $*^2=-1$ could equally well have been chosen. In the latter case one
would have to reverse the orientation of $V$ with respect to $\Sigma$, introducing a relative minus sign to the Wess-Zumino term and the
self-duality constraint would become ${\cal P}^M=-L^{MN}{\cal M}_{NP}*{\cal P}^P$.}
\begin{eqnarray}\label{eom}
\d*{\cal M}_{MN}{\cal P}^N+M_{NP}t_{MQ}{}^P{\cal P}^Q\wedge *{\cal P}^N+L_{MN}\d{\cal P}^N=0
\end{eqnarray}
Similar to the doubled torus sigma model constraint (\ref{constraint for doubled torus}), the self-duality constraint
\begin{equation}\label{constraint}
{\cal P}^M=L^{MN}{\cal M}_{NP}*{\cal P}^P
\end{equation}
can be consistently imposed. This constraint reduces the $2D$ degrees of freedom in $\mathbb{X}^I=(x^i,\tilde{x}_i)$ to the $D$ degrees of
freedom in the physical space-time coordinates $x^i$. In this way one can think of $D$ of the coordinates as auxiliary and the constraint
specifies how the auxiliary degrees of freedom depend on the physical degrees of freedom. Any two of
the Bianchi identity (\ref{bianchi}), Equation of motion (\ref{eom}) or constraint (\ref{constraint}) together determine the third. This
constraint is imposed in the quantum theory by gauging a left-acting maximally isotropic subgroup of $\cG$ as was done for the
Hamiltonian construction. Different choices of gauge group correspond to different ways of imposing the constraint. As stressed
earlier, there is no reason a-priori to think that different gaugings will give rise to equivalent theories. The gauged sigma model is given by
\begin{eqnarray}\label{b}
S=\frac{1}{4}\oint_{\Sigma}{\cal M}_{MN}\widehat{{\cal P}}^M\wedge *\widehat{{\cal P}}^N+\frac{1}{2}\oint_{\Sigma}L_{MN}{\cal P}^M\wedge{\cal
C}^N+\frac{1}{12}\int_Vt_{MNP}{\cal P}^M\wedge {\cal P}^N\wedge{\cal P}^P
\end{eqnarray}
which may be written as
$$
S=\frac{1}{4}\oint_{\Sigma}\d^2\sigma\,\,\sqrt{h}h^{\alpha\beta}\langle\widehat{{\cal P}}_{\alpha}|{\cal R}|\widehat{{\cal P}}_{\beta}\rangle +
\frac{1}{2}\oint_{\Sigma}\d^2\sigma\,\,\varepsilon^{\alpha\beta}\langle{\cal P}_{\alpha}|{\cal C}_{\beta}\rangle+
\frac{1}{12}\int_V\d^3\sigma'\,\,\varepsilon^{\alpha'\beta'\gamma'}\langle{\cal P}_{\alpha'}|[{\cal P}_{\beta'},{\cal P}_{\gamma'}]\rangle
$$
and is discussed further in Appendix E and \cite{HullRRE}.

\subsubsection{Derivation of the Poisson-Lie Map}

The details of this calculation are technically very similar to that of the chiral approach described above. The general results are
summarised here and further details of the calculation are given in the Appendix F.

A Manin triple is chosen such that an element of the double may be decomposed as $h=\tilde{g}g$ where $\tilde{g}\in
\widetilde{G}$ and $g\in G$. The sigma model on the group $G$ is recovered as the left coset $G=\cG/\widetilde{G}_L$ and, as for the chiral approach, the sigma model on the coset is given by gauging the left action $\widetilde{G}_L$ of the sigma model on $\cG$ or $\cX$. One-forms
$C=C_m\widetilde{T}^m$ which depend on both $\sigma$ and $\tau$ are introduced and gauging by minimal coupling is given by introducing the invariant
one-forms
\begin{equation}
\widehat{{\cal P}}=h^{-1}(\d+C)h
\end{equation}
The left-invariant one forms may be written as $\widehat{{\cal P}}^M=\widehat{\mathscr{P}}^N{\cal V}_N{}^M$ where
$$
\mathscr{P}^M=\left(%
\begin{array}{cc}
  r^m & \tilde{\ell}_m \\
\end{array}%
\right) \qquad  \widehat{\mathscr{P}}^M=\left(%
\begin{array}{cc}
  r^m & \widehat{\ell}_m \\
\end{array}%
\right)
$$
where $r^m=(\d gg^{-1})$ is the right-invariant form on $T^*G$, ${\ti \ell}_m=({\ti g}^{-1}\d{\ti g})$ is the left-invariant form on
$T^*\widetilde{G}$ and $\widehat{\ell}_m=({\ti g}^{-1}D{\ti g})$ is the gauge-invariant form on $T^*\widetilde{G}$ where the covariant
derivative $D=\d+C$ has been introduced. ${\cal V}$ is given by ${\cal O}$ in (\ref{adjoint action}). Define the $g$-dependent background
tensor ${\cal M}(\mathcal{F})$ as
$$
{\cal M}(\mathcal{F})={\cal V} {\cal M}(E){\cal V}^T
$$
where, as is discussed in Appendix B,
$$
{\cal F}_{mn}(g)=\left(A_p{}^m+E_{pq}\beta^{qm}\right)^{-1}E_{pt}(A^{-1})^t{}_n
$$
As demonstrated in in the Chiral (Hamiltonain) approach, the Wess-Zumino term may be written as
$$
S_{\text{wz}}=\frac{1}{2}\oint_{\Sigma}r^m\wedge \tilde{\ell}_m
$$
so the gauged action (\ref{gauged}) can then be written as
$$
S=\frac{1}{4}\oint_{\Sigma}{\cal M}_{MN}(\mathcal{F})\widehat{\mathscr{P}}^M\wedge*\widehat{\mathscr{P}}^N
+\frac{1}{2}\oint_{\Sigma}L_{MN}\mathscr{P}^M\wedge\mathscr{C}^N +\frac{1}{2}\oint_{\Sigma}r^m\wedge \tilde{\ell}_m
$$
where $\mathscr{C}={\ti g}^{-1}\mathcal{C}{\ti g}$. The action is written completely in terms of the two-dimensional world-sheet $\Sigma$ and so
can be expressed as a Lagrangian. Expanding the Lagrangian out and completing the square in $\widehat{\ell}_{m}$ gives
$$
S=\oint_{\Sigma}\d^2\sigma\,\,\mathcal{F}_{mn}r_-{}^mr_+{}^n+\frac{1}{2}\oint_{\Sigma}\d^2\sigma\,\,G^{mn}\lambda_{-m}\lambda_{+n}
$$
where
$$
\lambda_m=\widehat{\ell}_{m}-G_{mn}*r^n-B_{mn}r^n
$$
Using (\ref{right to left}) and integrating out the gauge fields, the action takes the form
$$
S=\oint_{\Sigma}d^2\sigma\,\,\mathcal{E}_{mn}(g^{-1}\partial_-g)^m(g^{-1}\partial_+g)^n
$$
where $\mathcal{E}_{mn}$ is given by (\ref{E}). Using instead $\cG=g\tilde{g}$ where ${\ti g}\in\widetilde{G}$ and $g\in G$ and gauging the
left-acting $G_L$ symmetry of the sigma model, the sigma model on $\widetilde{G}=\cG/G_L$ can be shown to be given by the action
$$
S=\oint_{\Sigma}d^2\sigma\,\,\widetilde{\mathcal{E}}^{mn}({\ti g}^{-1}\partial_-{\ti g})_m({\ti g}^{-1}\partial_+{\ti g})_n
$$
with $\widetilde{\mathcal{E}}^{mn}$ given by (\ref{E dual}). The doubled action (\ref{b}) presented here correctly reproduces the
Poisson-Lie map and includes generalizations to backgrounds with non-trivial flux.

\section{M2-branes and M-theory}

One might also consider generlising the analysis presented here to M-theory. M2 branes do not share the same status of fundamental quanta that strings have, so it is unclear whether or not a generalization of the doubled formalism of strings to M2 branes is appropriate; however, such a construction may elucidate many of the issues which arise in the construction and interpretation of many non-geometric backgrounds. When the internal space is a torus, the doubled geometry comes from introducing coordinates, conjugate to the string winding modes, in addition to the standard coordinates of the spacetime, conjugate to the momentum modes. Extending this, by analogy, to M2 branes, one expects the an extended geometry constructed by the coordinates, conjugate to string momentum and winding modes, to be replaced by a geometry with coordinates conjugate to membrane momentum and wrapping modes \cite{Hull ``Flux compactifications of string theory on twisted tori'',Hull ``Flux compactifications of M-theory on twisted tori''} (see also \cite{Cederwall:2007je,Hull:Generalised geometry for M-theory}). U-duality would have a natural action on the geometry of this extended space as a sub-goup of the mapping class group.

For string theory on a Poisson-Lie background, the cover of the associated doubled geometry $\cG/\G$ is dictated by a Lie-algebra - the gauge algebra $\mathfrak{h}=T\cG$ of the supergravity. This gauge algebra is related to the generalized tangent bundle $T\oplus T^*$ of the spacetime. Given such a Lie group $G$ with Poisson structure, one can define a bracket on the cotangent bundle $T^*G$ using the Poisson bracket. One may then define a natural algebraic structure on $TG\oplus T^*G$. In the case where the group $G$ is a torus,  $TG$ is related to momentum modes of the string and $T^*G$ is related to the winding modes. This algebraic structure on $TG\oplus T^*G$ is isomorphic to the Lie
algebra $\mathfrak{g}$, which in this case is that of a Drinfel'd double. It would be interesting to investigate a natural generalization of this construction to Nambu-Lie groups - Lie groups with a generalization of a
Poisson bracket, called a Nambu bracket \cite{Nambu}, which allows one to define an algebraic structure on $TG\oplus\bigwedge^2T^*G$. If $G$ is a torus, $TG$ is related to momenta and it is tempting to now suggest that $\bigwedge^2T^*G$ is related to the spectrum of M2 wrapping modes.

\begin{appendix}

\section{Semi-direct product example}

If the Poisson-structure is trivial $[\ell,\ell']^*=0$, then the natural bracket on $TG\oplus T^*G$ is the
Courant bracket
$$
[K+\ell,K'+\ell']=[K,K']+{\cal L}_{K}\ell'-{\cal L}_{K'}\ell-\frac{1}{2}\d\left(\iota_K\ell'-\iota_{K'}\ell\right)
$$
where $K\in TG$ and $\ell\in T^*G$. The Lie derivative\footnote{Consider the left-invariant one-form $\ell=g^{-1}dg$. The right action on
$\ell$, generated by the left-invariant vector field $\xi=\xi^mK_m$ is
$$
R_{\xi}:g\mapsto g\xi   \qquad  R_{\xi}\ell=\xi^{-1}d\xi+\xi^{-1}\ell\xi
$$
Infinitesimally $\xi=1+\varepsilon^mT_m$, then
$$
\delta_{\varepsilon}\ell^m=d\varepsilon^m+f_{np}{}^m\ell^n\varepsilon^p
$$
 This is precisely the action of the Lie derivative along the left-invariant vector field $\varepsilon=\varepsilon^mK_m$
$$
 {\cal L}_{\varepsilon}\ell^m=d\iota_{\varepsilon}\ell^m+\iota_\varepsilon d\ell^m=d\varepsilon^m+f_{np}{}^m\ell^n\varepsilon^p
$$
 so that ${\cal L}_K\ell=ad^*_K\ell$ in this case.} ${\cal L}_K\ell$ generates the right action of the vector field $K$ on the left-invariant one-form
$\ell$, in other words it is the adjoint action of $G$ on $T^*G$. Using
$$
\d\ell^m+\frac{1}{2}f_{np}{}^m\ell^n\wedge\ell^p=0   \qquad  [K_m,K_n]=f_{mn}{}^pK_p
$$
the bracket may be written as
$$
\left[K_m,K_n\right]=f_{mn}{}^pK_p\qquad \left[\ell^m,K_n\right]=f_{np}{}^m\ell^p\qquad \left[\ell^m,\ell^n\right]=0
$$
As an example, consider the double $\cG=G\ltimes \R^D$ where $G$ is a $D$-dimensional group with Lie algebra
$$
[T_m,T_n]=f_{mn}{}^pT_p \qquad  [\widetilde{T}^m,T_n]=f_{np}{}^m\widetilde{T}^p \qquad  [\widetilde{T}^m,\widetilde{T}^n]=0
$$
is given in terms of the generators $t_m$ of $G$
$$
T_m=\left(%
\begin{array}{cc}
  t_m & 0 \\
  0 & t_m \\
\end{array}%
\right) \qquad  \widetilde{T}^m=\left(%
\begin{array}{cc}
  0 & h^{mn}t_n \\
  0 & 0 \\
\end{array}%
\right)
$$
where $[t_m,t_n]=f_{mn}{}^pt_p$ and $h^{mn}$ is an invariant metric of $G$ (e.g. Cartan-Killing if $G$ is semi-simple). For consideration of the
left-invariant objects a useful parameterisation of a general element of this group is
$$
h=\left(%
\begin{array}{cc}
  g & g\tilde{x}_mh^{mn}t_n \\
  0 & g \\
\end{array}%
\right) \qquad  h^{-1}=\left(%
\begin{array}{cc}
  g^{-1} & -\tilde{x}_mh^{mn}t_ng^{-1} \\
  0 & g^{-1} \\
\end{array}%
\right)
$$
where $h\in\cG$, $g\in G$ and $\tilde{x}_m$ are coordinates on the group $\widetilde{G}=\R^D$. Note that in this parameterisation the
coordinate $\tilde{X}$ is Lie-algebra valued\footnote{Actually, $\tilde{x}_m$ takes values in the Lie co-algebra.} i.e.
$\tilde{x}_i=\delta_i{}^m\tilde{x}_m$. The left-invariant forms ${\cal P}=h^{-1}\d h\in\mathfrak{h}$ are
$$
{\cal P}=\left(%
\begin{array}{cc}
  g^{-1}\d g & \d\tilde{x}-[\tilde{x},g^{-1}\d g] \\
  0 & g^{-1}\d g \\
\end{array}%
\right)
$$
or ${\cal P}=\ell^mT_m+Q_m\widetilde{T}^m$ where
$$
\ell^m=(g^{-1}\d g)^m    \qquad  Q_m=\d\tilde{x}_m+f_{mn}{}^p\tilde{x}_p\ell^n
$$
These forms satisfy the Bianchi identities
$$
d\ell^m+\frac{1}{2}f_{np}{}^m\ell^n\wedge \ell^p=0    \qquad  \d Q_m-f_{mn}{}^pQ_p\wedge \ell^n=0
$$
Dual to these one-forms are the generators of the right action $Z_m$ and $X^m$
$$
Z_m=(\ell^{-1})_m{}^i\partial_i-f_{mn}{}^p\tilde{x}_p\partial^n    \qquad  X^m=\partial^m
$$
This gives a non-trivial lift of the isometry group of $G$, generated by $K_m=(\ell^{-1})_m{}^i\partial_i$, to the double $h$. These
generators satisfy
$$
[Z_m,Z_n]=-f_{mn}{}^pZ_p \qquad  [X^m,Z_n]=f_{np}{}^mX^p \qquad  [X^m,X^n]=0
$$
These objects are invariant under the rigid left action given by
$$
g\sim hg    \qquad  \tilde{x}\sim \tilde{x}+g^{-1}\tilde{\alpha}g
$$
We see that if we restrict to the subspace invariant under $X^m$, parameterized by the coordinates $x^i$, $Z_m$ becomes
$$
Z_m|_{G}=K_m
$$

\section{Adjoint Actions and Background Tensors}

At the identity $e\in\cG$ we could define a tensor $E_{mn}$ such that
\begin{equation}
\mathscr{E}^+=Span\{T_m+E_{mn}(e)\widetilde{T}^n\}   \qquad  \mathscr{E}^-=Span\{T_m-E_{nm}(e)\widetilde{T}^n\}
\end{equation}
We can define a tensor ${\cal F}_{mn}(g)$ at generic points $g\in G\subset\cG$ such that ${\cal F}_{mn}(e)=E_{mn}$. This may be achieved
by acting with the adjoint action on the eigenspaces $\mathscr{E}^{\pm}$
\begin{equation}
g^{-1}\mathscr{E}^+g=Span\{T_m+{\cal F}_{mn}(g)\widetilde{T}^n\}   \qquad  g^{-1}\mathscr{E}^-g=Span\{T_m-{\cal F}_{nm}(g)\widetilde{T}^n\}
\end{equation}
To determine the explicit form of ${\cal F}_{mn}(g)$ consider
\begin{eqnarray}
g^{-1}\mathscr{E}^+g&=&Span\{g^{-1}T_mg+E_{mn}g^{-1}\widetilde{T}^ng\}\nonumber\\
&=&Span\{\left(A_m{}^p+E_{mn}\beta^{np}\right)T_p+E_{mn}(A^{-1})^n{}_p\widetilde{T}^p\}
\end{eqnarray}
where (\ref{adjoint action}) has been used. The overall normalization has no meaning and so we can divide out by the factor
$\left(A_m{}^p+E_{mn}(e)\beta^{np}\right)$ throughout and identify the coefficient in front of the $\widetilde{T}^m$ generator as
${\cal F}_{mn}(g)$, i.e.
\begin{equation}
{\cal F}_{mn}(g)=\left(A_p{}^m+E_{pq}\beta^{qm}\right)^{-1}E_{pt}(A^{-1})^t{}_n
\end{equation}
we define the symmetric and anti-symmetric parts of ${\cal F}(g)_{mn}$ as $G_{mn}(g)$ and $B_{mn}(g)$ respectively.  Using the fact that
$r^m{}_i=A_n{}^m(g)\ell^n{}_i$ we can write
\begin{eqnarray}\label{right to left}
{\cal F}_{mn}(g)r^m{}_ir^m{}_j&=&{\cal E}_{mn}(g)\ell^m{}_i\ell^m{}_j
\end{eqnarray}
where $({\cal E}^{-1})^{mn}(g)=E^{mn}+\pi^{mn}(g)$. Note that the open string theory sees a background composed of a
metric $G_{open}{}^{mn}=E^{(mn)}$ and a non-commutativity parameter $\theta^{mn}=E^{[mn]}+\pi^{mn}$. Similar arguments can be given for the
transport of $\mathscr{E}^{\pm}$ by the adjoint action of $\widetilde{G}$ where one defines a $\tilde{g}$-dependent background tensor
$\widetilde{\cal F}$ through
\begin{equation}
\tilde{g}^{-1}\mathscr{E}^+\tilde{g}=\text{Span}\{\widetilde{T}^m+\widetilde{{\cal F}}^{mn}(\tilde{g})T_n\}   \qquad
\tilde{g}^{-1}\mathscr{E}^-\tilde{g}=\text{Span}\{\widetilde{T}^m-\widetilde{{\cal F}}^{nm}(\tilde{g})T_n\}
\end{equation}
It is not hard to show that the tensor $\widetilde{{\cal F}}^{mn}(\tilde{g})$ is given by
\begin{equation}
\widetilde{{\cal
F}}^{mn}(\tilde{g})=\left(\widetilde{A}^p{}_m+\widetilde{E}^{pq}\widetilde{\beta}_{qm}\right)^{-1}\widetilde{E}^{pq}(\widetilde{A}^{-1})_q{}^n
\end{equation}
We can also define
\begin{eqnarray}
\widetilde{{\cal F}}^{mn}(\tilde{g})\tilde{r}_m{}^i\tilde{r}_n{}^j=\widetilde{{\cal E}}^{mn}(\ti g)\tilde{\ell}_m{}^i\tilde{\ell}_n{}^j
\end{eqnarray}
where $\widetilde{{\cal E}}(\ti g)=(E^{-1}+{\ti \pi})^{-1}$. The Poisson-Lie map takes us from a sigma model with Lagrangian
\begin{equation}
\mathscr{L}={\cal F}_{mn}(g)(\partial_-gg^{-1})^m(\partial_+gg^{-1})^n={\cal E}_{mn}(\tilde{g})(g^{-1}\partial_-g)^m(g^{-1}\partial_+g)^n
\end{equation}
where ${\cal E}(g)=\left(E^{-1}(e)+\pi\right)$ to one with Lagrangian
\begin{equation}
\mathscr{L}=\widetilde{{\cal F}}^{mn}(\tilde{g})(\partial_-\tilde{g}\tilde{g}^{-1})_m(\partial_+\tilde{g}\tilde{g}^{-1})_n =\widetilde{\cal
E}^{mn}(\tilde{g})(\tilde{g}^{-1}\partial_-\tilde{g})_m(\tilde{g}^{-1}\partial_+\tilde{g})_n
\end{equation}
where ${\cal E}(\ti g)=\left(E+\pi\right)^{-1}$. We see that in the case where $\cX=T^{2D}$, then $\pi={\ti \pi}=0$ and the Buscher rules
$\widetilde{{\cal E}}={\cal E}^{-1}$ are recovered.

\section{Metric and Product Structure on the Double}

In this appendix, the explicit form of the real structure ${\cal R}$ is given.
\begin{equation}
{\cal R}=|\mathscr{E}^{+m}\rangle \langle\mathscr{E}_m^+|+|\mathscr{E}^{m-}\rangle \langle\mathscr{E}^-{}_m|
\end{equation}
The identity is
\begin{equation}
\bid=|\mathscr{E}^{+m}\rangle \langle\mathscr{E}_m^+|-|\mathscr{E}^{-m}\rangle \langle\mathscr{E}^-{}_m|
\end{equation}
The matrix representation of the linear idempotent map ${\cal R}$ is given by
\begin{equation}
{\cal R}^M{}_N=\langle T^M|{\cal R}|T_N\rangle
\end{equation}
If we define the symmetric metric ${\cal M}_{MN}=L_{MP}{\cal R}^P{}_N$ then we can write
\begin{equation}
{\cal R}^M{}_N=L^{MP}\langle T_P|{\cal M}|T_N\rangle
\end{equation}
It is actually the metric ${\cal M}$ which plays a more fundamental role in the analysis. The matrix takes the form
\begin{equation}
{\cal M}_{MN}=\left(%
\begin{array}{cc}
  {\cal M}^{mn} & {\cal M}_m{}^n \\
  {\cal M}^m{}_n & {\cal M}_{mn} \\
\end{array}%
\right)
\end{equation}
It is not too hard to show that
\begin{eqnarray}
\langle T_m|\mathscr{E}_n^{\pm}\rangle=\pm \frac{1}{\sqrt{2}}E_{nm}    \qquad \langle
\widetilde{T}^m|\mathscr{E}_n^{\pm}\rangle=\frac{1}{\sqrt{2}}\delta^m{}_n
\end{eqnarray}
from which we can show
\begin{eqnarray}
{\cal M}_{mn}&=&\langle T_m|{\cal M}|T_n\rangle=\langle T_m|\mathscr{E}^{p+}\rangle \langle\mathscr{E}_p^+|T_n\rangle +\langle T_m|\mathscr{E}^{p-}\rangle \langle\mathscr{E}_p^-|T_n\rangle\nonumber\\
&=&E_{pm}g^{pq}E_{nq}=g_{mn}+B_{mp}g^{pq}B_{qn}
\end{eqnarray}
\begin{eqnarray}
{\cal M}^m{}_n&=&\langle \widetilde{T}^m|{\cal M}|T_n\rangle=\langle \widetilde{T}^m|\mathscr{E}^{p+}\rangle \langle\mathscr{E}_p^+|T_n\rangle +\langle \widetilde{T}^m|\mathscr{E}^{p-}\rangle\langle\mathscr{E}_p^-|T_n\rangle\nonumber\\
&=&g^{mp}E_{[pn]}=g^{mp}B_{pn}
\end{eqnarray}
\begin{eqnarray}
{\cal M}^{mn}&=&\langle \widetilde{T}^m|{\cal M}|\widetilde{T}^n\rangle=\langle \widetilde{T}^m|\mathscr{E}^{p+}\rangle
\langle\mathscr{E}_p^+|\widetilde{T}^n\rangle+\langle \widetilde{T}^m|\mathscr{E}_p^-\rangle\langle\mathscr{E}_q^-|\widetilde{T}^n\rangle=g^{mn}
\end{eqnarray}
so that
\begin{eqnarray}
{\cal M}_{MN}=\left(%
\begin{array}{cc}
  g_{mn}+B_{mp}g^{pq}B_{qn} & g^{np}B_{pm} \\
  g^{mp}B_{pn} & g^{mn} \\
\end{array}%
\right)
\end{eqnarray}
The real structure is given by ${\cal R}^M{}_N=L^{MP}{\cal M}_{PN}$
\begin{eqnarray}
{\cal R}^M{}_N=\left(%
\begin{array}{cc}
  g^{mp}B_{pn} & g_{mn}+B_{mp}g^{pq}B_{qn} \\
  g^{mn} & g^{mp}B_{pn} \\
\end{array}%
\right)
\end{eqnarray}

There is a corresponding reduction of the structure group in terms of the components of the generalised vectors on $TG\oplus T^*G$ representation. Define the basis
$$
|\mathscr{E}^+{}_m)=\frac{1}{\sqrt{2}}[|K_m)+E_{mn}|\ell^n)]   \qquad  |\mathscr{E}^-{}_m)=\frac{1}{\sqrt{2}}[|K_m)-E_{nm}|\ell^n)]
$$
We now consider $E$ to be the Span of $g\pm B:T^*G\rightarrow TG$. The explicit forms of the matrix product structure, which in this basis we
denote as $\mathscr{R}$, are
\begin{equation}
\mathscr{R}=|\mathscr{E}^{+m})(\mathscr{E}_m^+|+|\mathscr{E}^{m-})(\mathscr{E}^-{}_m|
\end{equation}
The identity is
\begin{equation}
\bid=|\mathscr{E}^{+m})(\mathscr{E}_m^+|-|\mathscr{E}^{-m})(\mathscr{E}^-{}_m|
\end{equation}
where $|\mathscr{E}^{\pm m})=g^{mn}|\mathscr{E}^{\pm}{}_n)$. Using the fact that $(K_m|\ell^n)=\delta_m{}^n$ it is not hard to show that
$\mathcal{M}_{MN}=({\cal Z}_M|\mathscr{R}|{\cal Z}_N)$ takes the same form as before. In other words, with a minor abuse of notation,
$$
\mathcal{M}_{MN}=({\cal Z}_M|\mathscr{R}|{\cal Z}_N)=\langle T_M|{\cal R}|T_N\rangle
$$

\section{Adjoint Transformation of the Doubled Metric}

In this appendix we show the adjoint action of $G$ (or alternatively $\widetilde{G}$) on the basis elements $T_M$ induces an $O(D,D)$
transformation on the metric $\mathcal{M}$. Let $\mathcal{M}_{MN}(g)=\langle g^{-1}T_Mg|\mathcal{R}|g^{-1}T_Ng\rangle$ then
\begin{eqnarray}
\mathcal{M}_{MN}(g)&=&\langle g^{-1}T_Mg|\mathcal{R}|g^{-1}T_Ng\rangle\nonumber\\
&=&\langle g^{-1}T_Mg|T^P\rangle\langle T_P|\mathcal{R}|T^Q\rangle
\langle T_Q|g^{-1}T_Ng\rangle\nonumber
\end{eqnarray}
Using $g^{-1}T_Mg=T_S\mathcal{O}^S{}_M(g)$ and $\langle T_P|\mathcal{R}|T^Q\rangle=\mathcal{M}_{PQ}(E)$
we can write this as
\begin{eqnarray}
\langle T_S|\mathcal{O}^S{}_M(g)|T^P\rangle \mathcal{M}_{PQ}(E)
\langle T_Q|\mathcal{O}_N{}^T(g)|T_T\rangle =\mathcal{O}_M{}^P(g)\mathcal{M}_{PQ}(E) \mathcal{O}^Q{}_M(g)\nonumber
\end{eqnarray}
so that
$$
\mathcal{M}_{MN}(g)=\mathcal{O}_M{}^P(g)\mathcal{M}_{PQ}(E) \mathcal{O}^Q{}_M(g)
$$
as required.

\section{H-twisted Drinfel'd double}

As a simple example, consider the $H$-twisted Drinfel'd double $\mathfrak{h}_H$. It was explicitly shown how the sigma
model on the group $G$ (or the twisted torus $G/\G_G$) with $G_L$-invariant $H$-flux was recovered from the doubled sigma model
(\ref{Klimcik action}) by gauging the subgroup $\widetilde{G}_L$. Here, the same result is recovered by starting with the new sigma model
(\ref{fibre action}) and gauging the same $\widetilde{G}_L$ subgroup. The left-invariant one-forms may be written as
$$
{\cal P}^M=\left(%
\begin{array}{cc}
  P^m & Q_m \\
\end{array}%
\right)
$$
where the Bianchi identities (\ref{bianchi}) in this case are
\begin{equation}\label{example}
\d P^m+\frac{1}{2}f_{np}{}^mP^n\wedge P^p=0   \qquad  \d Q_m- f_{mn}{}^pQ_p\wedge P^n-\frac{1}{2}H_{mnp}P^n\wedge P^p=0
\end{equation}
These one-forms are equivalent to the (left-invariant) right-acting gauge algebra
$$
[Z_m,Z_n]=f_{mn}{}^pZ_p+H_{mnp}X^p   \qquad  [Z_m,X^n]=-f_{mp}{}^nX^p   \qquad  [X^m,X^n]=0
$$
The right-invariant gauge group is obtained by changing the signs of the structure constants and exchanging the left-invariant generators $Z_m$
and $X^m$ with the right-invariant generators $\widetilde{Z}_m$ and $\widetilde{X}^m$ respectively. The key observation is that
$[\widetilde{X}^m,\widetilde{X}^n]=0$ and so the $\widetilde{X}^m$ generate a maximally isotropic subgroup $\widetilde{G}_L\subset \cG_L$
which can be gauged. Gauging the symmetry generated by $\widetilde{X}^m$ requires the introduction of the gauge fields
$C=C_m\widetilde{T}^m$ by minimal coupling
$$
{\cal P}=h^{-1} \d h\rightarrow \widehat{{\cal P}}=h^{-1} \left(\d+C\right)h
$$
It is useful to define ${\cal C}=h^{-1} Ch$, so that
$$
\widehat{{\cal P}}^M={\cal P}^M+\mathcal{C}^M \qquad  {\cal C}^M=\left(h^{-1}Ch\right)^M
$$
In the current example (\ref{example}) one can show that ${\cal C}={\cal C}_m\widetilde{T}^m$, i.e. there is no $\mathcal{C}^mT_m$ component and the minimal
coupling may be written simply as
$$
{\cal P}^M\rightarrow \widehat{{\cal P}}^M={\cal P}^M+{\cal C}^M    \qquad  \Leftrightarrow \qquad Q_m\rightarrow \widehat{Q}_m=Q_m+C_m
$$
As discussed in \cite{Hull and Spence}, the term
$$
\frac{1}{2}L_{MN}{\cal P}^M\wedge \mathcal{C}^N
$$
must also be added to give the gauged action
\begin{eqnarray}\label{fibre action}
S=\frac{1}{4}\oint_{\Sigma}{\cal M}_{MN}\widehat{{\cal P}}^M\wedge *\widehat{{\cal P}}^N+\frac{1}{2}\oint_{\Sigma}L_{MN}{\cal P}^M\wedge{\cal
C}^N+\frac{1}{12}\int_Vt_{MNP}{\cal P}^M\wedge {\cal P}^N\wedge{\cal P}^P
\end{eqnarray}
which may be written as
$$
S=\frac{1}{4}\oint_{\Sigma}\d^2\sigma\,\,\sqrt{h}h^{\alpha\beta}\langle\widehat{{\cal P}}_{\alpha}|{\cal R}|\widehat{{\cal P}}_{\beta}\rangle +
\frac{1}{2}\oint_{\Sigma}\d^2\sigma\,\,\varepsilon^{\alpha\beta}\langle{\cal P}_{\alpha}|{\cal C}_{\beta}\rangle+
\frac{1}{12}\int_V\d^3\sigma'\,\,\varepsilon^{\alpha'\beta'\gamma'}\langle{\cal P}_{\alpha'}|[{\cal P}_{\beta'},{\cal P}_{\gamma'}]\rangle
$$
This is not strictly independent of the choice of Lie-algebra basis as the definition of ${\cal C}$ depends on a choice of which maximally
isotropic subgroup of $\cG$ is gauged. Using the Bianchi identities (\ref{example}), the Wess-Zumino term in (\ref{gauged}) is
\begin{eqnarray}\label{WZW}
S_{wz}&=&\frac{1}{4}\int_Vf_{np}{}^mQ_m\wedge P^n\wedge P^p-\frac{1}{12}\int_VH_{mnp}P^m\wedge P^n\wedge P^p\nonumber\\
&=&\frac{1}{2}\int_{\Sigma}P^m\wedge Q_m+\frac{1}{6}\int_VH_{mnp}P^m\wedge P^n\wedge P^p
\end{eqnarray}
The fact that $P^m\wedge Q_m$ is globally defined has been used to write the two-dimensional term. Expanding the action (\ref{fibre action})
using (\ref{WZW}) and (\ref{M matrix}) and then completing the square in $C_m$, the doubled action may be written
$$
S=\oint_{\Sigma}\left(\frac{1}{2}g_{mn}P^m\wedge *P^n+\frac{1}{2}B_{mn}P^m\wedge P^n+\frac{1}{4}g^{mn}\lambda_m\wedge
*\lambda_n\right)+\int_V\frac{1}{6}H_{mnp}P^m\wedge P^n\wedge P^p\nonumber
$$
where
$$
\lambda_m=Q_m+\mathcal{C}_m-g_{mn}*P^n-B_{mn}P^n
$$
The $\mathcal{C}_m$ can be integrated out to give a theory whose target space is a twisted torus with $G_L$-invariant $H$-flux, reproducing the
result of the last section.

\section{Recovering the Poisson-Lie Map}

This Appendix shows how the Poisson-Lie map is recovered from the doubled formalism. We assume the double may be decomposed (at least close to the identity) as $h=\tilde{g}g$ where $\tilde{g}\in \cG/G_L$ and $g\in G$. The
left-invariant one forms may be written as
\begin{equation}
{\cal P}=g^{-1}\left(r+\tilde{\ell}\right)g=g^{-1}\mathscr{P}g
\end{equation}
where
\begin{eqnarray}
\mathscr{P}^M=\left(%
\begin{array}{cc}
  r^m & \tilde{\ell}_m \\
\end{array}%
\right)\qquad    r=\d gg^{-1}   \qquad  \tilde{\ell}=\tilde{h}^{-1}\d\tilde{h}
\end{eqnarray}
with Bianchi identities (\ref{structure equations}). Note that ${\cal P}$ takes values in the cotangent bundle of $\cG$, whereas $\mathscr{P}$ takes values in the cotangent bundle of $G\times \widetilde{G}$. Choosing a basis for the generators we
have $r=r^mT_m$ and $\tilde{\ell}=\tilde{\ell}_m\widetilde{T}^m$ so that
\begin{equation}
{\cal P}=r^m\left(g^{-1}T_mg\right)+\tilde{\ell}_m\left(g^{-1}\widetilde{T}^mg\right)=\mathscr{P}^M(g^{-1}T_Mg)
\end{equation}
Recalling the adjoint action of $G$ on $T\cG$ (\ref{adjoint action}) we can write ${\cal P}^M=\mathscr{P}^N{\cal O}_N{}^M$ where ${\cal
O}$ is given by (\ref{adjoint action}). Define the $g$-dependent background tensor ${\cal M}(\mathcal{E})$ as
\begin{equation}
{\cal M}(\mathcal{F})={\cal O} {\cal M}(E){\cal O}^T
\end{equation}
As demonstrated in section three, the Wess-Zumino term may be written as
$$
S_{wz}=\frac{1}{2}\oint_{\Sigma}r^m\wedge \tilde{\ell}_m
$$
which is the generalisation of the symplectic form used in the abelian case. The action now only depends on the worldsheet $\Sigma$ and can be
written in terms of the Lagrangian density
$$
\mathscr{L} =\frac{1}{4}{\cal M}_{MN}(\mathcal{F})\mathscr{P}^M\wedge*\mathscr{P}^N+\frac{1}{2}r^m\wedge \tilde{\ell}_m
$$

\subsection{Gauging the Left Action}

Gauging by minimal coupling is given by introducing the invariant one-forms
\begin{equation}
\widehat{{\cal P}}=h^{-1}(\d+C)h
\end{equation}
Under infinitesimal variations from the left $\delta C=-\d\alpha-[\alpha,C]$ and $\delta\widehat{{\cal P}}=0$, where the
parameter may be written as $\alpha=\alpha^mT_m+\alpha_m\widetilde{T}^m$. Choosing first the polarisation $h=\tilde{g}g$
\begin{equation}
\widehat{{\cal P}}=g^{-1}\widehat{\mathscr{P}}g
\end{equation}
where
\begin{equation}
\widehat{\mathscr{P}}=\left(%
\begin{array}{cc}
  r & \widehat{\ell} \\
\end{array}%
\right) \qquad r=\d gg^{-1}   \qquad  \widehat{\ell}=\tilde{g}^{-1}(\d+C)\tilde{g}
\end{equation}
We are therefore considering a gauging of the left action $\delta\tilde{g}=\alpha\tilde{g}, \delta g=0$. Under this action the fields transform as
\begin{eqnarray}
\delta r=0\qquad \delta \tilde{\ell}=\tilde{g}^{-1}\d\alpha \tilde{g}\qquad \delta C_m=-\d\alpha_m-c_m{}^{np}\alpha_pC_n
\end{eqnarray}
Consider the term
\begin{equation}
\frac{1}{2}L_{MN}{\cal P}^M\wedge {\cal C}^N=\frac{1}{2}\langle{\cal P}|{\cal C}\rangle
=\frac{1}{2}\langle\mathscr{P}|\mathscr{C}\rangle=\frac{1}{2}L_{MN}\mathscr{P}^M\wedge\mathscr{C}^N
\end{equation}
where $\mathscr{C}={\ti g}^{-1}\mathcal{C}{\ti g}$. The gauged action can then be written as
$$
S=\frac{1}{4}\oint_{\Sigma}{\cal M}_{MN}(\mathcal{F})\widehat{\mathscr{P}}^M\wedge*\widehat{\mathscr{P}}^N
+\frac{1}{2}\oint_{\Sigma}L_{MN}\mathscr{P}^M\wedge\mathscr{C}^N +\frac{1}{2}\oint_{\Sigma}r^m\wedge \tilde{\ell}_m
$$
Note that since ${\ti g}^{-1}\widetilde{T}^m{\ti g}\in \widetilde{G}$ we have $\mathscr{C}^m=0$ and
$$
\frac{1}{2}\oint_{\Sigma}L_{MN}\mathscr{P}^M\wedge \mathscr{C}^N+\frac{1}{2}\oint_{\Sigma}r^m\wedge{\ti \ell}_m=\frac{1}{2}\oint_{\Sigma}r^m\wedge\hat{\ell}_m
$$
Expanding out the action
\begin{eqnarray}
S&=&\frac{1}{4}\oint_{\Sigma}G^{mn}\widehat{\ell}_{m}\wedge
*\widehat{\ell}_{n}+\frac{1}{2}\oint_{\Sigma}G^{mp}B_{pn}\widehat{\ell}_{m}\wedge\ast
r^n +\frac{1}{4}\oint_{\Sigma}(G_{mn}+B_{mp}G^{pq}B_{qn})r^m\wedge\ast r^n \nonumber\\
&&+\oint_{\Sigma}\frac{1}{2}r^m\wedge \widehat{\ell}_{m}
\end{eqnarray}
Completing the square in $\widehat{\ell}_{m}$ gives
\begin{eqnarray}
S&=&\frac{1}{4}\oint_{\Sigma}G^{mn}\lambda_{m}\wedge *\lambda_{n} +\frac{1}{2}\oint_{\Sigma}G_{mn}r^m\wedge\ast r^n
+\frac{1}{2}\oint_{\Sigma}B_{mn}r^m\wedge r^n
\end{eqnarray}
where
\begin{eqnarray}
\lambda_m=\widehat{\ell}_{m}-G_{mn}*r^n-B_{mn}r^n
\end{eqnarray}
Using the result (\ref{right to left}) this can be written in the standard way (\ref{lagrangian}).

\subsection{Dual Polarization}

Consider instead $h=g\tilde{g}$ where ${\ti g}\in\widetilde{G}$ and $g\in \cG/\widetilde{G}_L$ so that
\begin{eqnarray}
{\cal P}=\tilde{g}^{-1}\widetilde{\mathscr{P}}\tilde{g}    \qquad    {\cal P}^M=\widetilde{\mathscr P}^N \widetilde{\cal O}_N{}^M
\end{eqnarray}
where $\widetilde{\cal O}$ is given by (\ref{adjoint action})
\begin{eqnarray}
\widetilde{\mathscr{P}}^M=(%
\begin{array}{cc}
  \tilde{r}^m & \ell_m \\
\end{array}%
)
\end{eqnarray}
$\tilde{r}=\d{\ti g}{\ti g}^{-1}$ and $\ell=\d gg^{-1}$. Note that
\begin{eqnarray}
\mathcal{M}(E)_{MN}\widehat{\mathcal{P}}^M\wedge*\widehat{\mathcal{P}}^N
&=&\mathcal{M}_{MN}(E+\tilde{\pi})\widetilde{\mathscr{P}}^M\wedge*\widetilde{\mathscr{P}}^N\nonumber\\
&=&\mathcal{M}_{MN}(\widetilde{\mathcal{E}}^{-1})\widetilde{\mathscr{P}}^M\wedge*\widetilde{\mathscr{P}}^N
=\mathcal{M}^{MN}(\widetilde{\mathcal{E}})\widetilde{\mathscr{P}}_M\wedge*\widetilde{\mathscr{P}}_N\nonumber
\end{eqnarray}
where $\widetilde{\mathscr{P}}_M=L_{MN}\widetilde{\mathscr{P}}^N$. The action in the dual polarisation becomes
\begin{eqnarray}
S&=&\frac{1}{4}\oint_{\Sigma}\mathcal{M}^{MN}(\widetilde{\mathcal{E}})\widetilde{\mathscr{P}}_M\wedge*\widetilde{\mathscr{P}}_N
+\frac{1}{2}\oint_{\Sigma}\ell_m\wedge\tilde{r}^m
\end{eqnarray}
In this polarisation, $h=g\tilde{g}$
\begin{equation}
\widehat{{\cal P}}=\tilde{g}^{-1}\left(\d\tilde{g}\tilde{g}^{-1}+g^{-1}(\d+C)g\right)\tilde{g}
\end{equation}
Under this action the fields transform as
\begin{eqnarray}
\delta \ell=0\qquad \delta \tilde{r}=g^{-1}\d\alpha g\qquad \delta C^m=-\d\alpha^m-f_{np}{}^m\alpha^pC^n
\end{eqnarray}
Proceeding as above, the gauged action can be shown to reproduce (\ref{Lagrangian dual}).

\end{appendix}

\begin{footnotesize}

\end{footnotesize}

\end{document}